\documentclass[12pt]{article}
\newcommand{\Comment}[1]{{}}
\usepackage{amsfonts,amsthm,amsmath,amssymb,slashed}
\usepackage[textwidth = 430 pt, textheight = 630 pt]{geometry}

\Comment{\usepackage{color}
\definecolor{MyDarkBlue}{rgb}{0.15,0.15,0.45}
\usepackage[linktocpage=true]{hyperref}
\hypersetup{
colorlinks=true,
citecolor=MyDarkBlue,\bibliography{myref}
linkcolor=MyDarkBlue,
urlcolor=MyDarkBlue,
pdfauthor={Thiago Araujo and Horatiu Nastase},
pdftitle={Non-Abelian T-duality for nonrelativistic holographic duals},
pdfsubject={hep-th}
}

\usepackage[numbers,sort&compress]{natbib}
\usepackage{hypernat}}
\usepackage{graphicx}

\newcommand\ignore[1]{}
\def\one{{\,\hbox{1\kern-.8mm l}}}

\def\a{\alpha}\def\b{\beta}
\def\g{\gamma}
\def\G{\Gamma}

\def\r{\rho}
\def\s{\sigma}
\def\S{\Sigma}
\def\L{\Lambda}

\def\d{\partial}

\newcommand{\Cset}{{\,\,{{{^{_{\pmb{\mid}}}}\kern-.45em{\mathrm C}}}}}

\newcommand{\be}{\begin{equation}}
\newcommand{\bea}{\begin{eqnarray}}

\newcommand{\ee}{\end{equation}}
\newcommand{\eea}{\end{eqnarray}}

\newcommand{\nn}{\nonumber}

\usepackage{color}
\newcommand{\bse}{\begin{subequations}}
\newcommand{\ese}{\end{subequations}}
\usepackage{graphicx}
\usepackage{caption}
\usepackage{subcaption}

\begin{document}

\renewcommand{\thefootnote}{\fnsymbol{footnote}}

\makeatletter
\@addtoreset{equation}{section}
\makeatother
\renewcommand{\theequation}{\thesection.\arabic{equation}}

\rightline{}
\rightline{}
   \vspace{1.8truecm}


\vspace{10pt}


\begin{center}
{\LARGE \bf{\sc Non-Abelian T-duality for nonrelativistic holographic duals}}
\end{center}
 \vspace{1truecm}
\thispagestyle{empty} \centerline{
 {\large \bf {\sc Thiago R. Araujo${}^{a,}$}}\footnote{E-mail address: \Comment{\href{mailto:taraujo@ift.unesp.br}}{\tt taraujo@ift.unesp.br}}
{\bf{\sc and}}
   {\large \bf {\sc Horatiu Nastase${}^{a,}$}}\footnote{E-mail address: \Comment{\href{mailto:nastase@ift.unesp.br}}{\tt
    nastase@ift.unesp.br}}                                                        }

\vspace{1cm}

\vspace{.8cm}
\centerline{{\it ${}^a$
Instituto de F\'{i}sica Te\'{o}rica, UNESP-Universidade Estadual Paulista}} \centerline{{\it
R. Dr. Bento T. Ferraz 271, Bl. II, Sao Paulo 01140-070, SP, Brazil}}

\vspace{1.0truecm}

\thispagestyle{empty}

\centerline{\sc Abstract}

\vspace{.4truecm}

\begin{center}
\begin{minipage}[c]{380pt}
{\noindent We find new type II backgrounds with non-relativistic symmetries via non-Abelian T-duality. First we consider geometries with Galilean
symmetries in type IIA, which have been identified as non-relativistic generalizations of the ABJM background and massive IIA supergravities. 
We then consider the non-Abelian T-duality transformation on the backgrounds with Lifshitz symmetry constructed by Donos and Gauntlett. Using 
gauge/gravity duality we study aspects of the field theory dual to these backgrounds.
}
\end{minipage}
\end{center}

\today

\vspace{.5cm}

\setcounter{page}{0}
\setcounter{tocdepth}{2}

\newpage

\tableofcontents
\renewcommand{\thefootnote}{\arabic{footnote}}
\setcounter{footnote}{0}

\linespread{1.1}
\parskip 4pt


\section{Introduction}

Gauge/gravity duality \cite{Maldacena:1997re,Gubser:1998bc,Witten:1998qj} (for a standard review, see \cite{Aharony:1999ti}) has had a lot of success
describing gauge theories, when derived from the decoupling limit of a system of branes. 

But a lot of the recent interest has been instead focused on applications to condensed matter physics, specifically in the study of strongly coupled 
systems described by relativistic and also nonrelativistic field theories. Since gauge/gravity duality relates strong coupling in field theory to weak 
coupling in gravity (and vice versa), we can analyze models that are otherwise very difficult to analyze. Since however in these AdS/CMT cases 
we usually have no decoupled system of branes, only a phenomenological construction of a gravity dual, we have usually less control over the 
construction. One degree of control is obtained by analyzing symmetry.

The symmetries of the field theory are realized geometrically as isometries in the gravity dual side. In the best known example, type IIB string 
theory on $AdS_5 \times S^5$ vs. ${\cal N}=4$ SYM, the isometry group of the gravity dual is $SO(4,2)\times SO(6)$, matching with the one
of a four dimensional conformal field theory with ${\cal N}=4$ supersymmetries \cite{Maldacena:1997re}.
While the $SO(4,2)$ symmetry of the anti-de Sitter space is reinterpreted as the conformal group in $3+1$ dimensions, the group 
$SO(6)\simeq SU(4)$ can be identified with the R-symmetry group of the conformal theory. 

In the case of strongly coupled condensed matter systems, we have a variety of numerical and theoretical tools from statistical physics and 
quantum field theory \cite{anderson2008basic,sachdev2011quantum}, but they usually are hard to use. In the case of relativistic systems, the tools of 
AdS/CFT are applied in the usual way. But in the case of systems with  non-relativistic spacetime symmetries, we first define the non-relativistic 
algebra and then we try to realize it geometrically 
\cite{Nishida:2007pj,Son:2008ye,Balasubramanian:2008dm,Herzog:2008wg,Maldacena:2008wh,Adams:2008wt}.

There are two symmetry algebras that are relevant in the nonrelativistic case (see \cite{Hartnoll:2009sz} and references therein). The first one, 
known as Lifshitz algebra, contains the generators for rotations $\{M_{ij}\}$, translations $\{P_i\}$, time translations $\{H\}$ and dilatations $D$, 
satisfying the standard commutation relations for $\{M_{ij}, P_i, H\}$ together with 
\bse
\be
[D, M_{ij}]=0\; , \quad [D, P_{i}]=iP_j\; , \quad [D, H]=i zH\; ,
\ee
and in \cite{Kachru:2008yh} the geometric realization of the above symmetry (which has been embedded in string theory in 
\cite{Balasubramanian:2010uk, Donos:2010tu}) was defined by the gravity dual
\be 
ds^2=L^2\left(-\frac{dt^2}{r^{2z}}+\frac{dx^i dx^i}{r^2}+\frac{dr^2}{r^2} \right).
\ee
\ese
As we can see, for $z=1$ we recover Anti-de Sitter space. 

A second relevant algebra is the conformal Galilean algebra which contains, besides the generators for rotations $\{M_{ij}\}$, translations 
$\{P_i\}$, time translations $\{H\}$ and dilatations $D$, also the  "Galilean boosts", generated by $K_i$, with nontrivial commutators
\bse
\be
\begin{split}
[M_{ij}, K_k]=i(\delta_{ik}K_j-\delta_{jk}K_i)&\; , \quad [ P_{i}, K_j]=-i\delta_{ij}N\; , \quad [K_i, H]=-i P_i\; ,\\
&\quad [D, K_i]=i(1-z)K_i\;,
\end{split}
\ee
In the special case $z=2$, the algebra is called the Schr\"{o}dinger algebra. 
Here $N$ is the number operator, which counts the number of particles with a given mass $m$, and in general has  has only one 
nontrivial commutation relation, $[D,N]=i(2-z)N$, but in the $z=2$ (Schr\"{o}dinger) case we see that it becomes a central charge. 
Curiously, it is not possible to arrange the $D$-dimensional Schr\"odinger algebra as an isometry in $(D+1)$-dimensions, but in 
\cite{Son:2008ye, Balasubramanian:2008dm} it was realized that we can write a gravity dual as  a $(D+2)$-dimensional space, 
with metric
\be 
ds^2=L^2\left(-\frac{dt^2}{r^{2z}}+\frac{-2dtd\xi + dx^i dx^i}{r^2}+\frac{dr^2}{r^2} \right)\; .
\ee
\ese

Obtaining nonrelativistic gravity duals in string theory turns out to be difficult (see \cite{Herzog:2008wg, Maldacena:2008wh, Adams:2008wt, Balasubramanian:2010uk, Donos:2010tu, Bobev:2009mw, Donos:2009en, Donos:2009xc, Ooguri:2009cv, Donos:2009zf, Singh:2010rt, Jeong:2009aa, Gregory:2010gx}).\footnote{See \cite{Ko:2015rha} for the embedding of nonrelativistic string backgrounds via the use of abelian T-duality in the context of double field theory.} 
In relativistic cases, several different techniques have been employed in order to generate supergravity solutions, see 
\cite{Gauntlett:2004hs, Gran:2005wu, Lazaroiu:2013fg} for recent developments.
One particularly interesting solution generating technique which has been applied extensively is T-duality. In the usual case,  T-duality 
relates strings in a background with a compact direction, $S^1$ of radius $R$, with a background with an $S^1$ of radius $\a'/R$. 
The physical spectrum of a string in the geometry is invariant under this transformation, see e.g. 
\cite{Buscher:1987qj, Rocek:1991ps, Giveon:1991jj, Alvarez:1993qi, Alvarez:1994dn}.

This usual duality (on $S^1$) is abelian ($U(1)$ group), but a non-Abelian generalization for the group $SU(2)$, called non-Abelian T-duality,
was introduced in  \cite{delaOssa:1992vc} and became an issue of recent interest 
\cite{Giveon:1993ai, Gasperini:1993nz, Alvarez:1994np, Sfetsos:2010uq, Lozano:2011kb, Itsios:2013wd, Kelekci:2014ima}. This non-abelian T-duality 
(NATD) transformation has been used successfully as a solution generating technique 
\cite{Lozano:2012au, Itsios:2012zv, Lozano:2013oma, Caceres:2014uoa, Lozano:2014ata, Araujo:2015npa, Bea:2015fja, Macpherson:2013zba, Barranco:2013fza, Gaillard:2013vsa, Kooner:2014cqa, Lozano:2015bra, Lozano:2015cra}, 
although some issues concerning global properties of the dual manifold remain.

Considering the difficulties in constructing string theory gravity duals with non-relativistic symmetries, in this paper we consider NATD 
of known gravity dual solutions. In section 2, we apply this technique to the solutions with conformal Galilean symmetry constructed 
in \cite{Singh:2010rt}, and in section 3 to the solutions with Lifshitz symmetries constructed in \cite{Donos:2010tu}.

In order to define the dual field theory, in section 4 we start by calculating the quantized Page charges  of the spaces constructed in the section 2 and 3. 
In particular, we compare the charges of the Galilean solution constructed in section 2 with the charges calculated in \cite{Lozano:2014ata}. 
We then define and study holographic Wilson loops in these backgrounds, and in section 5 we conclude.

\section{Galilean Solutions}\label{galilean}

In this section, we first give a short review of the solutions of \cite{Singh:2010rt}, which are nonrelativistic generalizations of the gravity dual to
ABJM \cite{Aharony:2008ug} in type IIA string theory. We then perform non-Abelian T-duality on them, obtaining new type IIB backgrounds. 
We consider the solutions in  \cite{Singh:2010rt} because they have the nontrivial $z=3$, even though we don't know much about their 
holographic dual field theory.

\subsection{Galilean type-IIA solution and its NATD}

The Galilean nonrelativistic solution of type IIA string theory of \cite{Singh:2010rt} has string frame metric\footnote{Using the fact that the constant 
$\b$ is arbitrary, we make the transformation $\b\to \frac{1}{\sqrt{2}}\b$, compared with \cite{Singh:2010rt}. Also, remember that 
$g_{\mu\nu}^{str}= e^{\frac{4}{D-2}\phi}g_{\mu\nu}^E$.} 
\begin{equation}
ds_{IIA}^2=\underbrace{\frac{ R^2}{4}\left(-\frac{\beta^2(dx^+)^2}{z^6}+\frac{dy^2+dz^2-2dx^+dx^-}{z^2}\right)}_{ds^2_{Gal}}
+ R^2\ ds_{\mathbb{CP}^3}^2 \label{metricgal}
\end{equation}
where $R^2=\sqrt{ \tilde{R}^3/k}$ and the Fubini-Study metric for $\mathbb{CP}^3$ is (see, e.g. \cite{Nishioka:2008gz,Cvetic:2000yp})
\begin{align}
ds_{\mathbb{CP}^3}^2&=d\zeta^2+\frac{1}{4}\cos^2\zeta(d\theta_1^2
+\sin^2\theta_1 d\phi_1^2)+\frac{1}{4}\sin^2\zeta(d\theta_2^2+\sin^2\theta_2 d\phi_2^2)\nonumber\\
&+\frac{1}{4}\sin^2\zeta \cos^2\zeta (d\psi+\cos\theta_1 d\phi_1+\cos\theta_2 d\phi_2)^2,\nonumber \\
&=d\zeta^2+\frac{1}{4}\cos^2\zeta ds_1^2+\frac{1}{4}\sin^2\zeta (\tau_1^2+\tau_2^2)
+\frac{1}{4}\sin^2\zeta \cos^2\zeta \left(\tau_3+\cos\theta_1 d\phi_1\right)^2\; .\label{cp3}
\end{align} 
Here $ds_1^2=d\theta_1+\sin^2\theta_1 d\phi_1^2$, $\zeta\in [0,\pi/ 2]$, $\theta_i \in [0,\pi]$, 
$\phi_i \in [0,2\pi]$, $\psi\in [0,4\pi]$ and $\tau_i$ are the Maurer-Cartan forms for the group $SU(2)$, namely
\bea
\tau_1&=&-\sin\psi d\theta_2+\cos\psi\sin\theta_2 d\phi_2\nonumber\\
\tau_2&=&\cos\psi d\theta_2+\sin\psi\sin\theta_2 d\phi_2\\ \label{taui} 
\tau_3&=&d\psi +\cos\theta_2 d\phi_2\nonumber\; ,
\eea
with $d \tau_i=\frac{1}{2}\epsilon_{ijk}\tau_j \wedge \tau_k$. Considering the equation (\ref{spmetric}), we see that
\bse
\begin{equation}
ds_7^2=ds_{Gal}^2+ R^2 d\zeta^2+\frac{ R^2}{4}\cos^2\zeta ds_1^2\; ,
\end{equation}
\begin{equation}
\sum_{i=1}^3 e^{2C_i}(\tau_i+A^i)^2=\frac{R^2}{4}\sin^2\zeta(\tau_1^2+\tau_2^2)+\frac{R^2}{4}\sin^2\zeta \cos^2\zeta(\tau_3+\cos\theta_1 d\phi_1)^2\; ,
\end{equation}
that is,
\begin{equation}
\begin{split}
e^{C_1}&=e^{C_2}=\frac{R}{2}\sin \zeta \equiv \b_1^{1/2}\; , \quad e^{C_3}=\frac{R}{2}\sin\zeta\cos \zeta\equiv \b_2^{1/2}\; ,\\
\quad A^1&=A^2=0\; , \quad A^3=\cos \theta_1 d\phi_1\; .
\end{split}
\end{equation}
\ese
We define the vielbeins associated to the Galilean metric
\begin{subequations}
\begin{equation}
ds^2_{Gal}=-\mathfrak{e}^+\mathfrak{e}^+ + \mathfrak{e}^-\mathfrak{e}^- +\mathfrak{e}^y\mathfrak{e}^y+\mathfrak{e}^z\mathfrak{e}^z\; ,
\end{equation}
as
\begin{equation}
\begin{split}
\mathfrak{e}^+&=\frac{R\beta}{2}\left(\frac{1}{z^3} dx^+ + \frac{z}{\b^2}dx^-\right)\, ,\quad \mathfrak{e}^-=\frac{Rz}{2\b}dx^-\\
\mathfrak{e}^y&= \frac{R}{2z} dy\, , \quad \mathfrak{e}^z= \frac{R}{2z} dz\; .
\end{split}
\end{equation}
\end{subequations}
This solution is also supplemented with the following fields
\bea
e^\phi&=&\frac{R}{k}\; , \quad  B=\frac{\beta}{\sqrt{2}} \frac{R^2 p}{z^4} dx^+\wedge d y\; , \\ 
C_{(1)}&=&\frac{\beta}{\sqrt{2}}\frac{q k}{z^3}dx^+ + 2 k \omega\; , \cr
dC_{(3)}&=&\frac{3R^2 k}{8z^4}dx^+\wedge dx^-\wedge dy\wedge dz=\frac{6k}{R^2} e^+\wedge e^-\wedge e^\wedge e^z\; ,
\eea
where $q=2p=\frac{1}{\sqrt{2}}$, $\mathcal{J}=d\omega$ is a K\"{a}hler $2$-form on 
$\mathbb{CP}^3$ and the level $k$ is the quantum of $dC_{(1)}$ on $\mathbb{CP}^1\in\mathbb{CP}^3$ , that is 
\be
\int_{\mathbb{CP}^1}dC_{(1)}=2\pi k\; .\label{quant.01}
\ee
Considering that on the $\mathbb{CP}^1(\theta_1,\phi_1)$ with $\zeta=0$ we have (e.g. \cite{Lozano:2014ata})
\begin{equation}
\begin{split}
\omega&=-\frac{1}{4}\sin^2\zeta \left(\tau_3+\cos\theta_1 d\phi_1 \right)+\frac{1}{4}\cos\theta_1 d\phi_1\\
&=-\frac{1}{2}\tan \zeta\; \mathfrak{E}_3+\frac{1}{4}\cos\theta_1 d\phi_1 \; ,
\end{split}
\end{equation}
we have
\begin{equation}
\mathcal{J}=\mathfrak{E}_3\wedge\mathfrak{E}_\zeta+\mathfrak{E}_\phi\wedge\mathfrak{E}_\theta-
\mathfrak{E}_1\wedge\mathfrak{E}_2
\end{equation}
where we have defined the following vielbeins with relation to the metric $ds_{\mathbb{CP}^3}^2$ in (\ref{cp3})
\begin{equation}
\begin{split}
\mathfrak{E}_\zeta&=d\zeta\; ,\quad \mathfrak{E}_\theta=\frac{1}{2}\cos\zeta d\theta_1\; ,\quad 
\mathfrak{E}_\phi=\frac{1}{2}\cos \zeta \sin\theta_1 d\phi_1\\
\mathfrak{E}_1 &= \frac{1}{2}\sin\zeta \tau_1\; ,\quad \mathfrak{E}_2 = \frac{1}{2}\sin\zeta \tau_2\; , \quad
\mathfrak{E}_3 =\frac{1}{2}\sin\zeta \cos\zeta (\tau_3+\cos\theta_1 d\phi_1)\; .
\end{split}
\end{equation}
With these definitions we can easily see that $vol (\mathbb{CP}^3)=\frac{1}{3!}\mathcal{J}\wedge \mathcal{J}\wedge \mathcal{J}$, that is
\begin{equation}
\begin{split}
vol (\mathbb{CP}^3)&=\mathfrak{E}_\zeta\wedge \mathfrak{E}_\phi\wedge 
\mathfrak{E}_\theta\wedge \mathfrak{E}_1 \wedge \mathfrak{E}_2\wedge \mathfrak{E}_3\\
&=\frac{1}{32}\cos^3\zeta\sin^3\zeta \sin \theta_1\sin \theta_2 d\zeta \wedge d\theta_1\wedge d\phi_1 \wedge d\theta_2\wedge d\phi_2 \wedge d \psi\; .
\end{split}
\end{equation}
Therefore, using the quantization of the Page charge
\be
\frac{1}{(2\pi \a'^{1/2})^{7-p}}\int_{\Sigma^{8-p}}\mathcal{F}_{8-p}=\mathcal{Q}_{Dp}\in \mathbb{Z}\; ,
\ee
where $\mathcal{F}=F \wedge e^{-B}$, for some cycle $\Sigma^{8-p}$, we can see that
\be 
\frac{1}{(2\pi \a'^{1/2})^5}\int_{\mathbb{CP}^3}\ast \mathcal{F}_4=N_{D2}\in \mathbb{Z}\; \Rightarrow \; R^4=\frac{32\pi^2 \a'^{5/2}}{k}N_{D2}\; .\label{quant.02}
\ee

The vielbeins with relation to the metric (\ref{metricgal}) for the internal space are defined as
\begin{subequations}
\begin{align}
\mathfrak{e}_\zeta&=R d\zeta\; ,\quad \mathfrak{e}_\theta=\frac{R}{2}\cos\zeta d\theta_1\; ,\quad 
\mathfrak{e}_\phi=\frac{R}{2}\cos \zeta \sin\theta_1 d\phi_1\; ,\\
\mathfrak{e}_1 &= \frac{R}{2}\sin\zeta \tau_1\equiv \beta^{1/2}_1 \tau_1\; ,\quad \mathfrak{e}_2 = \frac{R}{2}\sin\zeta \tau_2 \equiv \beta^{1/2}_1 \tau_2\; ,\\
\mathfrak{e}_3 &=\frac{ R}{2} \sin\zeta \cos\zeta \left(\tau_3+\cos\theta_1 d\phi_1\right)\equiv \beta^{1/2}_2 (\tau_3+A_3)\; .
\end{align}
\end{subequations}
The relativistic limit of this solution, that is $AdS_4\times \mathbb{CP}^3$, can be recovered by setting $\b\to 0$.

\subsubsection{Nonabelian T-dual of the Galilean background}

Now we want to perform a T-duality transformation \cite{Itsios:2013wd, Kelekci:2014ima, Itsios:2012zv, Gevorgyan:2013xka}  
with respect to the $SU(2)$ isometry. We construct the matrix $M_{ij}$, defined by $M_{ij}=g_{ij}+b_{ij}+\a'\epsilon_{ijk}\hat v_k$, obtaining
(since $b_{ij}=0$),
\begin{subequations}
\begin{equation}
M=\begin{pmatrix}
\beta_1 & \a'\hat{v}_3 & -\a'\hat{v}_2 \\ 
-\a'\hat{v}_3 & \beta_1 & \a'\hat{v}_1 \\ 
\a'\hat{v}_2 & -\a'\hat{v}_1 & \beta_2
\end{pmatrix}.
\end{equation} 
We consider a gauge where $\theta_2=\phi_2=v_2=0$, so that $\hat{v}=(\cos \psi v_1, \sin\psi v_1, v_3)$, where $\psi \in [0,2\pi]$. We can make 
connections the gauge choice in \cite{Lozano:2014ata} by making the transformation $(v_1=\rho \sin \chi, v_3=\rho \cos \chi) $ with $\chi \in[0,\pi]$ 
and the range of the coordinate $\rho$ is not yet determined, but we argue that $\rho \in[n\pi, (n+1)\pi)$ as in 
\cite{Lozano:2014ata, Macpherson:2014eza}, (see \cite{Itsios:2013wd} for other possible gauge choices). 

Therefore, the matrix $M$ in this gauge is
\begin{equation}
M=\begin{pmatrix}
\beta_1 & \a'v_3 & -\a'\sin\psi v_1 \\ 
-\a' v_3 & \beta_1 & \a' \cos\psi v_1 \\ 
\a'\sin\psi v_1 & -\a' \cos\psi v_1 & \beta_2
\end{pmatrix}.
\end{equation}
\end{subequations}
The dilaton in the dual theory is given by
\begin{equation}
\widehat{\phi}=\phi-\frac{1}{2}\ln \left(\frac{\Delta}{\a'^3}\right)\; \Rightarrow\; e^{\widehat{\phi}}=\frac{R \a'^{3/2}}{k \Delta^{1/2}}\ ,
\end{equation}
where $\Delta\equiv\det M=[(\beta_1^2+\a'^2v_3^2)\beta_2+\a'^2v_1^2\beta_1]$. 


Using the results of Appendix \ref{review}, the dual metric becomes
\begin{subequations}
\begin{equation}
d\hat{s}_{IIB}^2=ds_7^2+\frac{1}{\Delta}d\Sigma^2
\end{equation}
where
\begin{align}
d\Sigma^2&=(z_1 Dz_1+z_2 Dz_2+z_3 Dz_3)^2+e^{2(C_2+C_3)}Dz_1^2+e^{2(C_1+C_3)}Dz_2^2+e^{2(C_1+C_2)}Dz_3^2 \nonumber\\
&=\a'^2 v_1^2 \beta_1\beta_2 \hat{\eta}^2+\a'^2\left\{(\beta_1\beta_2+\a'^2 v_1^2)dv_1^2+(\beta_1^2+\a'^2v_3^2)dv_3^2+2\a'^2 v_1v_3dv_1dv_3\right\} ,
\end{align}
with $\hat{\eta}=d\psi+\cos\theta_1 d\phi_1$.
\end{subequations}
Here we have used that $z_a=\a' \hat{v}_a$ and
\begin{equation}
\frac{1}{\a^\prime}Dz_1= d\hat{v}_1-\hat{v}_2 A^3\; , \quad \frac{1}{\a^\prime}Dz_2= d\hat{v}_2+\hat{v}_1 A^3\; ,\quad 
\frac{1}{\a^\prime}Dz_3= d\hat{v}_3\; .
\end{equation}

The dual vielbeins are
\begin{equation}
\hat{\mathfrak{e}}_a=e^{C_a}\Delta^{-1}\left[z_a z_b Dz_b+e^{2\sum_{b\neq a}C_b}Dz_a+\epsilon_{abc}z_b e^{2C_b}Dz_c \right]\; ,
\end{equation}
such that
\bse
\begin{align}
\hat{\mathfrak{e}}'_1&\equiv\cos \psi \hat{\mathfrak{e}}_1+\sin\psi \hat{\mathfrak{e}}_2=\frac{\a'\b_1^{1/2}}{\Delta}\left[(\b_1\b_2
+\a'^2v_1^2)dv_1+\a'^2v_1v_3 dv_3-\a' \b_2 v_1 v_3\hat{\eta}  \right]\\
\hat{\mathfrak{e}}'_2&\equiv-\sin \psi \hat{\mathfrak{e}}_1+\cos\psi \hat{\mathfrak{e}}_2=
\frac{\a'\b_1^{1/2}}{\Delta}\left[\a'\b_2 v_3 dv_1-\a'\b_1 v_1 dv_3+\b_1 \b_2 v_1 \hat{\eta}  \right]\\
\hat{\mathfrak{e}}_3&=\frac{\a'\b_2^{1/2}}{\Delta}\left[\a'^2 v_1 v_3 dv_1+(\b_1^2+\a'^2v_3^2)dv_3+\a'\b_1v_1^2 \hat{\eta} \right]\; .
\end{align}
\ese

The Kalb-Ramond field is given by
\bse
\begin{align}
\widehat{B}&=\frac{\beta}{\sqrt{2}} \frac{R^2 p}{z^4} dx^+\wedge d y\nonumber\\
&-\frac{1}{\Delta}\left\{e^{2C_1}z_1 Dz_2\wedge Dz_3+ e^{2C_2}z_2 Dz_3\wedge Dz_1+e^{2C_3}z_3 Dz_1\wedge Dz_2\right\}-Dz_3\wedge A^3\nn\\
&= \frac{\beta}{\sqrt{2}} \frac{R^2 p}{z^4} dx^+\wedge d y-\frac{\a'^3v_1v_3\b_2}{\Delta}dv_1\wedge \hat{\eta}
+\frac{\a'(\a'^2 v_1^2\b_1-\Delta)}{\Delta}dv_3\wedge (\cos\theta_1 d\phi_1)\nn\\
&+\frac{\a'^3v_1^2\b_1}{\Delta}dv_3\wedge d\psi  \nn \\
&=\frac{\beta}{\sqrt{2}} \frac{R^2 p}{z^4} dx^+\wedge d y-\underbrace{\frac{\a'\beta_2}{\Delta}\left(\a'^2 v_1 v_3 dv_1
+(\a'^2 v_3^2+\b_1^2)dv_3\right)}_{=\sqrt{\b_2}\; \hat{\mathfrak{e}}^3\wedge \hat{\eta}}\wedge \hat{\eta}+\underbrace{\a' 
dv_3\wedge d\psi}_{(\textit{closed form})}\nn\\
\end{align}
or, using the spherical coordinates $(v_1,v_3)=(\rho \sin\chi, \rho \cos\chi)$,
\begin{align}
\widehat{B}&=\frac{\beta}{\sqrt{2}} \frac{R^2 p}{z^4} dx^+\wedge d y-\frac{\a'\beta_2}{\Delta}\left((\b_1^2+\a'^2\r^2)\cos\chi d\r
-\b_1^2\r\sin\chi d\chi \right)\wedge \hat{\eta} \nn\\
&=\frac{\beta}{\sqrt{2}} \frac{R^2 p}{z^4} dx^+\wedge d y - \a'\frac{R^2\sin^2\zeta}{64\Delta}\left\{
R^4\rho \cos^2\zeta\sin^4\zeta \cos\theta_1\sin \chi d\phi_1\wedge d\chi \phantom{\frac{1}{1}}
\right.\nn\\
&\phantom{\frac{1}{1}}+\cos^2\zeta \cos\theta_1 \cos\chi (R^4\sin^4\zeta+16\a'^2\rho^2)d\rho\wedge d\phi_1\\
&\left.\phantom{\frac{1}{1}}-\rho R^4\sin^4\zeta \cos^2\zeta \sin\chi d\chi\wedge d\psi+(R^4\sin^4\zeta+16\a'^2\rho^2)\cos^2\zeta 
\cos\chi d\rho \wedge d\psi\right\}\nn
\end{align}
 After a gauge transformation, we can write the $B$-field as
\begin{align}
\widehat{B}&=\frac{\beta}{\sqrt{2}} \frac{R^2 p}{z^4} dx^+ \wedge d y- \a'\frac{R^2\sin^2\zeta}{64\Delta}\left\{
R^4\rho \cos^2\zeta\sin^4\zeta \cos\theta_1\sin \chi d\phi_1\wedge d\chi \phantom{\frac{1}{1}}
\right.\nn\\
&\phantom{\frac{1}{1}.}+\cos^2\zeta \cos\theta_1 \cos\chi (R^4\sin^4\zeta+16\a'^2\rho^2)d\rho\wedge d\phi_1\\
&\left.\phantom{\frac{1}{1}}+16\a'^2\rho^2\left[\rho (\sin^2\chi+\cos^2\chi\cos^2\zeta)\sin\chi d\chi\wedge d\psi 
-\sin^2\chi\sin^2\zeta \cos\chi d\rho \wedge d\psi\right]\right\}\nn\\
&\phantom{\frac{1}{1}.}-\a'd(\rho \cos\chi d\psi)\nn\; ,
\end{align}
\ese
with the term on the last line being a pure gauge contribution. The $\widehat{B}$-field at the $2$-cycle defined 
by $\tilde{S}^2=(\phi_1=const., x^+=const., y=const.; \chi, \psi)$ is
\be
\left. \phantom{\frac{1}{1}}\widehat{B}\right|_{\tilde{S}^2}=-\a' \rho \sin \chi d\chi\wedge d\psi\; ,
\ee
where we also have used that $\lim_{\zeta\to 0}\frac{\Delta}{\sin^2\zeta}= \frac{R^2 \rho^2}{4}\a'^2$. Large gauge transformations 
are defined such that the holonomy of $\widehat{B}$ satisfies
\be 
b=\frac{1}{4\pi^2 \a'}\left| \int_{\tilde{S}^2}\widehat{B} \right| \in [n,n+1),
\ee
which justifies our choice $\rho\in [n\pi,(n+1)\pi)$. 

In order to find the dual R--R fields, it is convenient to write the RR-fields before the T-duality as
\begin{subequations}
\begin{align}
F_{2}&=dC_{(1)}=-12\frac{q k}{R^2}\mathfrak{e}^z\wedge \mathfrak{e}^t +\frac{2k}{R^2}(\mathfrak{e}_3\wedge \mathfrak{e}_\zeta
+\mathfrak{e}_\phi\wedge \mathfrak{e}_\theta-\mathfrak{e}_1\wedge \mathfrak{e}_2)\nonumber\\
&=\frac{2k}{R^2}\left(6q\; \mathfrak{e}^t\wedge \mathfrak{e}^z+\mathfrak{e}^\phi\wedge\mathfrak{e}^\theta\right)
-\frac{2k \b_2^{1/2}}{R^2}\mathfrak{e}_\zeta\wedge (\tau^3+A_3)-\frac{2k\b_1}{R^2}\tau_1\wedge \tau_2
\nonumber\\
&\equiv G_2+J_1^3\wedge (\tau^3+A_3)+K_0^3\tau_1\wedge \tau_2
\end{align}
\begin{align}
F_4&=dC_{(3)}-H\wedge C_{(1)}\nonumber\\
&=\frac{2 k}{R^2}\mathfrak{e}^t\wedge \mathfrak{e}^y\wedge \mathfrak{e}^z\wedge \left(3\mathfrak{e}^- + 8p 
\frac{\cot\theta_1}{\cos\zeta}\mathfrak{e}_\phi\right)- \frac{16 p k}{R^2}\tan\zeta\; \mathfrak{e}^t\wedge \mathfrak{e}^y
\wedge \mathfrak{e}^z\wedge \mathfrak{e}_3 \nonumber\\
&=\frac{2 k}{R^2}\mathfrak{e}^t\wedge \mathfrak{e}^y\wedge \mathfrak{e}^z\wedge \left(3\mathfrak{e}^- 
+  \frac{8p\cot\theta_1}{\cos\zeta}\mathfrak{e}_\phi\right)- \frac{16 p k \b_2^{1/2}}{R^2}\tan\zeta\; \mathfrak{e}^t\wedge 
\mathfrak{e}^y\wedge \mathfrak{e}^z\wedge  (\tau^3+A_3)\nonumber\\
&\equiv G_4+L_3^3\wedge (\tau^3+A_3)\; ,
\end{align}
\end{subequations}
with $\mathfrak{e}^t=(\mathfrak{e}^+-\mathfrak{e}^-)/\sqrt{2}$, $F_6=-\ast F_4$ and $F_8=\ast F_2$. Using equations (\ref{rra}--\ref{rre}) 
we have that the dual fields in the RR sector are given by
\begin{subequations}
\bea
\a^{\prime 3/2}\widehat{F}_{1}&=& -\a' v_3\; J_1^3-K_0^3\b_2^{1/2}\hat{\mathfrak{e}}_3+\a' K_0^3\b_1^{-1/2}v_1 
\left(\cos \psi \hat{\mathfrak{e}}_2-\sin\psi \hat{\mathfrak{e}}_1\right)\nn\\
	&=&-\a' v_3 J_1^3-\a' K_0^3 dv_3
\eea
\bea
\a^{\prime 3/2}\widehat{F}_{3}&=&\b_1\b_2^{1/2}\ast_7 G_4+\a' G_2\wedge \left[v_1 \b_1^{1/2}(\cos\psi\hat{\mathfrak{e}}_1
+\sin\psi \hat{\mathfrak{e}}_2 )+v_3 \b_2^{1/2} \hat{\mathfrak{e}}_3\right]\nn\\
&+&J_1^3\wedge\left[-\b_1 \hat{\mathfrak{e}}_1\wedge \hat{\mathfrak{e}}_2+\frac{\a'\b_1^{1/2}v_1}{\b_2^{1/2}}\hat{\mathfrak{e}}_3
\wedge\left(\cos\psi \hat{\mathfrak{e}}_1+\sin\psi \hat{\mathfrak{e}}_2 \right) \right]\cr
&+&\frac{\a'v_3 K_0^3\b_2^{1/2}}{\b_1}
\hat{\mathfrak{e}}_1\wedge \hat{\mathfrak{e}}_2\wedge \hat{\mathfrak{e}}_3\nn\\
&=&\b_1\b_2^{1/2}\ast_7 G_4+\a'^2 G_2\wedge\left( v_1 dv_1+v_3 dv_3\right)-\frac{\a'^2 v_1}{\Delta}J_1^3\wedge 
\left[(\a'^2v_1^2\b_1+\b_1^2\b_2) dv_1 \right.\nn\\	
&+&\left.v_1v_3\a'^2\b_1 dv_3\right]\wedge \hat{\eta}-\frac{\a'^4 v_3v_1}{\Delta}K_0^3\b_2 dv_1\wedge dv_3\wedge \hat{\eta}
\eea
\bea
\a^{\prime 3/2}\widehat{F}_{5}&=&(1+\ast)\left\{\a'G_4\wedge \left[v_1 \b_1^{1/2}(\cos\psi\hat{\mathfrak{e}}_1
+\sin\psi \hat{\mathfrak{e}}_2 )+v_3 \b_2^{1/2} \hat{\mathfrak{e}}_3\right]
\phantom{\frac{\b_2^{1/2}k}{R^2}}\right.\nn\\
&+&\left.\b_1\b_2^{1/2} G_2\wedge \hat{\mathfrak{e}}_1\wedge \hat{\mathfrak{e}}_2\wedge \hat{\mathfrak{e}}_3\right.\cr
&+&\left. L_3^3\wedge\left[-\b_1 \hat{\mathfrak{e}}_1\wedge \hat{\mathfrak{e}}_2+\frac{\a'\b_1^{1/2}v_1}{\b_2^{1/2}}
\hat{\mathfrak{e}}_3\wedge\left(\cos\psi \hat{\mathfrak{e}}_1+\sin\psi \hat{\mathfrak{e}}_2 \right)\right]
\right\}\nn\\
&=&-\b_1 \b_2^{1/2}\ast_7 G_2+\a'G_4\wedge \left[v_1 \b_1^{1/2}\hat{\mathfrak{e}}_1'+v_3 \b_2^{1/2} \hat{\mathfrak{e}}_3\right]\cr
&+&\ast_7 L^3_3\wedge\left[-\b_1 \hat{\mathfrak{e}}_3+ \frac{\a'\b_1^{1/2}v_1}{\b_2^{1/2}}\hat{\mathfrak{e}}_2'  \right]
+ L_3^3\wedge\left[-\b_1 \hat{\mathfrak{e}}_1\wedge \hat{\mathfrak{e}}_2+\frac{\a'\b_1^{1/2}v_1}{\b_2^{1/2}}\hat{\mathfrak{e}}_3\wedge 
\hat{\mathfrak{e}}_1'\right]\cr
&&-\a'\ast_7 G_4\wedge \left[v_1\b_1^{1/2}\hat{\mathfrak{e}}_2'\wedge \hat{\mathfrak{e}}_3
+v_3\b_2^{1/2}\hat{\mathfrak{e}}_1'\wedge \hat{\mathfrak{e}}_2' \right]\nn\\
&+&\b_1\b_2^{1/2} G_2\wedge \hat{\mathfrak{e}}_1\wedge \hat{\mathfrak{e}}_2\wedge \hat{\mathfrak{e}}_3\nn\\
&=&-\b_1 \b_2^{1/2}\ast_7 G_2+\a'^2 G_4\wedge \left[v_1 dv_1+v_3 dv_2\right]-\frac{\a'\b_1}{\b_2^{1/2}}(\ast_7 L^3_3)\wedge dv_3\nn\\
&-&\frac{v_1\a'^2}{\Delta} L_3^3\wedge\left[(\a'^2 v_1^2\b_1+\b_1^2\b_2)dv_1+\a'^2 v_1v_3\b_1 dv_3\right]\wedge \hat{\eta}\\
&-&\frac{\a'^3v_1\b_1\b_2^{1/2}}{\Delta}\ast_7 G_4\wedge \left[v_3\b_2 dv_1-v_1\b_1dv_3 \right]\wedge\hat{\eta}\cr
&-&\frac{\a'^3 v_1\b_1^2\b_2}{\Delta} G_2\wedge  dv_1\wedge dv_3\wedge \hat{\eta}\nn
\eea
or, using spherical coordinates,
\begin{align}
&\a^{\prime 3/2}\widehat{F}_{1}= \frac{k}{2}\left(-\rho \sin^2\zeta \sin\chi d\chi+\sin^2\zeta \cos\chi d\rho-\rho\sin2\zeta \cos\chi d\zeta \right)\\
&\a^{\prime 3/2}\widehat{F}_3=\frac{kR}{4}\sin^3\zeta \cos\zeta\left(3 \mathfrak{e}^\zeta\wedge \mathfrak{e}^\theta\wedge \mathfrak{e}^\phi
+\frac{8p\cot\theta_1}{\cos\zeta} \mathfrak{e}^-\wedge \mathfrak{e}^\zeta\wedge \mathfrak{e}^\theta\right)\nn\\
&+\frac{2k\a'^2}{R^2}\rho\left(6q\mathfrak{e}^t\wedge \mathfrak{e}^z +\mathfrak{e}^\phi\wedge \mathfrak{e}^\theta\right)\wedge d\rho
+\frac{\a'^4 R^2}{4\Delta}\sin^2\zeta \cos^2\zeta \rho^3\cos \chi \sin \chi d\rho \wedge d\chi \wedge \hat{\eta}\nn\\
&+\frac{\a'^2kR^2}{4\Delta}\rho\sin^3\zeta\cos\zeta\sin\chi\;  d\zeta\wedge\left[ \frac{R^4}{16}\sin^4\zeta\cos^2\zeta d(\rho\sin\chi)
+  \a'^2 \rho ^2\; \sin\chi\;  d\rho  \right]\wedge \hat{\eta}\\ \label{hatfthree}
&\a^{\prime 3/2}\widehat{F}_5=(1+\ast)\left\{\frac{2k\a'^2}{R^2}\rho\;  \mathfrak{e}^t\wedge \mathfrak{e}^y \wedge \mathfrak{e}^z\wedge 
\left( 3 \mathfrak{e}^- + \frac{8 p\cot\theta_1}{\cos\zeta}\mathfrak{e}^\phi\right)\wedge d\rho \right.\nn\\
&+\frac{\a'^3 k R^4 }{ 32 \Delta } \sin^6\zeta \cos^2\zeta\; \rho^2\sin\chi\left(6q\; \mathfrak{e}^t\wedge \mathfrak{e}^z
+\mathfrak{e}^\phi\wedge \mathfrak{e}^\theta\right)\wedge d\rho\wedge d\chi\wedge \hat{\eta}\nn\\
&\left.+\frac{2pkR\a'^2}{\Delta}\sin^4\zeta\; \rho \sin\chi \; \mathfrak{e}^t\wedge \mathfrak{e}^y\wedge \mathfrak{e}^z\wedge\right.\cr
&\left.\wedge\left[ \frac{R^4}{16}\sin^4\zeta\cos^2\zeta d(\rho\sin\chi)+  \a'^2 \rho ^2\; \sin\chi\;  d\rho  \right]\wedge \hat{\eta}\right\}\; .
\end{align}
\end{subequations}


\subsection{Galilean solution in massive type-IIA and its NATD}

We also have the following background, in the string frame  \cite{Singh:2010rt}
\begin{subequations}
\begin{equation}
ds_{mIIA}^2=\underbrace{a_0\left(-\frac{\beta^2(dx^+)^2}{z^6}+\frac{dy^2+dz^2-2dx^+dx^-}{z^2}\right)}_{ds^2_{mGal}}
+\frac{5a_0}{2}ds_{\mathbb{CP}^3}^2\label{mmetric}
\end{equation}
where $a_0\, f^{2}=2$, $L^2=a_0 e^{-\phi/2}$, $ f^{5/2}\,L^2=2\, m_0^{1/2}$ and $m_0$ is the Romans' mass. In this background we also have the 
nontrivial fields
\begin{equation}
\begin{split}
e^{2\phi}&=\frac{f^2}{m_0^2}\; , \quad  B= \frac{2\beta}{f^2}\frac{1}{z^4} dx^+\wedge dy\; ,\\
C_{(1)}&=\frac{2\sqrt{5}\beta m_0}{3f^2}\frac{1}{z^3}dx^+\; , \quad d C_{(3)}=4\frac{\sqrt{5} m_0}{f^4 z^4}dx^+\wedge dx^-\wedge dy\wedge dz\; .
\end{split}
\end{equation}
\end{subequations}
Note that the solution is similar to the one in the previous subsection, but there are subtle differences.
This solution does not preserve any supersymmetry even in the relativistic case $\b=0$. 
We write the metric (\ref{mmetric}) in a similar manner to the previous subsection, as
\bse
\begin{equation}
ds^2_{mIIA}=d\tilde{s}^2_7+\sum_{i=1}^3 e^{2\tilde{C}_i}(\s^i+A^i)^2
\end{equation}
where
\begin{equation} 
d\tilde{s}^2_7=\, ds^2_{mGal}+\frac{5a_0}{2}\left(d\zeta^2 +\frac{1}{4}\cos^2\zeta ds_1^2\right)
\end{equation}
and
\begin{equation}
\sum_{i=1}^3 e^{2\tilde{C}_i}(\s^i +A^i)^2=\frac{5a_0}{8}\left[\sin^2\zeta(\tau_1^2+\tau_2^2)+\sin^2\zeta \cos^2\zeta(\tau_3+\cos\theta_1 d\phi_1)^2\right]\;,
\end{equation}
such that
\begin{equation}
\begin{split}
e^{\tilde{C}_1}&=e^{\tilde{C}_2}=\frac{\sqrt{5a_0}}{2\sqrt{2}}\sin \zeta \equiv \tilde{\b}_1^{1/2}\; , \quad e^{\tilde{C}_3}
=\frac{\sqrt{5a_0}}{2\sqrt{2}}\sin\zeta\cos \zeta\equiv \tilde{\b}_2^{1/2}\; ,\\
\quad A^1&=A^2=0\; , \quad A^3=\cos \theta_1 d\phi_1\; .
\end{split}
\end{equation}
\ese

\subsubsection{Nonabelian T-dual of the massive Galilean background}

The dual dilaton is given by
\begin{equation}
\widehat{\phi}=\phi-\frac{1}{2}\ln \left(\frac{\tilde{\Delta}}{\a'^3}\right)\ ,
\end{equation}
where $\tilde{\Delta}=[(\tilde{\beta}_1^2+\a'^2v_3^2)\tilde{\beta}_2+\a'^2v_1^2\tilde{\beta}_1]$. 
The nonabelian T-dual metric is
\begin{subequations}
\begin{equation}
d\hat{s}_{mIIB}^2=ds_7^2+\frac{1}{\tilde{\Delta}}d\tilde{\Sigma}^2
\end{equation}
where
\begin{align}
d\tilde{\Sigma}^2&=(z_1 Dz_1+z_2 Dz_2+z_3 Dz_3)^2+e^{2(C_2+C_3)}Dz_1^2+e^{2(C_1+C_3)}Dz_2^2+e^{2(C_1+C_2)}Dz_3^2 \nonumber\\
&=\a'^2 v_1^2 \tilde{\beta}_1\tilde{\beta}_2 \hat{\eta}^2+\a'^2\left\{(\tilde{\beta}_1\tilde{\beta}_2
+\a'^2 v_1^2)dv_1^2+(\tilde{\beta}_1^2+\a'^2v_3^2)dv_3^2+2\a'^2 v_1v_3dv_1dv_3\right\} ,
\end{align}
with $\hat{\eta}=d\psi+\cos\theta_1 d\phi_1$\; .
\end{subequations}
Here we have used that
\begin{equation}
\frac{1}{\a'}Dz_1= d\hat{v}_1-\hat{v}_2 A^3\; , \quad \frac{1}{\a'}Dz_2= d\hat{v}_2+\hat{v}_1 A^3\; ,\quad 
\frac{1}{\a'}Dz_3= d\hat{v}_3\; .
\end{equation}
The dual Kalb-Ramond field is given by
\begin{align}
\widehat{B}&= \frac{2\b}{f^2 z^4} dx^+\wedge d y\nonumber\\
&-\frac{1}{\tilde{\Delta}}\left\{e^{2\tilde{C}_1}z_1 Dz_2\wedge Dz_3+ e^{2\tilde{C}_2}z_2 Dz_3\wedge Dz_1
+e^{2\tilde{C}_3}z_3 Dz_1\wedge Dz_2\right\}-Dz_3\wedge A^3\nn\\
&=  \frac{2\b}{f^2z^4} dx^+\wedge d y-\frac{\a'^3v_1v_3\tilde{\b}_2}{\tilde{\Delta}}dv_1\wedge \hat{\eta}
+\frac{\a'(\a'^2 v_1^2\tilde{\b}_1-\tilde{\Delta})}{\tilde{\Delta}}dv_3\wedge (\cos\theta_1 d\phi_1)\nn\\
&+\frac{\a'^3v_1^2\tilde{\b}_1}{\tilde{\Delta}}dv_3\wedge d\psi\nn\\
&= \frac{2\b}{f^2 z^4} dx^+\wedge d y-\frac{\a'\tilde{\beta}_2}{\tilde{\Delta}}\left(\a'^2 v_1 v_3 dv_1+(\a'^2 v_3^2
+\tilde{\beta}_1^2)dv_3\right)\wedge \hat{\eta}+{\rm closed}\; .
\end{align}
Given that the original R-R fields are
\bse
\begin{align}
F_0&=m_0\\
F_{2}&=dC_{(1)}+m_0B= m_0\, \mathfrak{e}^t_{m_0}\wedge\left(\sqrt{5}\; \mathfrak{e}^z_{m_0}+  \mathfrak{e}^y_{m_0}\right)\\
F_4&=dC_{(3)}- H\wedge C_{(1)}+\frac{m_0}{2}B\wedge B\nonumber\\
&=\sqrt{5}\; m_0\, \mathfrak{e}^t_{m_0}\wedge \mathfrak{e}^-_{m_0}\wedge \mathfrak{e}^y_{m_0}\wedge \mathfrak{e}^z_{m_0} \; ,
\end{align}\label{ffour}
\ese
where these vielbeins are related to the metric (\ref{mmetric}), it folllows that the T-dual fields are given by
\bse
\begin{align}
\a'^{3/2}\widehat{F}_1&=\a'm_0\left[v_1\tilde{\b}_1^{1/2}(\cos\psi \hat{\mathfrak{e}}^1+\sin\psi \hat{\mathfrak{e}}^2)
+v_3\tilde{\b}_2^{1/2} \hat{\mathfrak{e}}^3 \right]\nn\\
&=m_0 \a'^2 (v_1 dv_1+v_3dv_3)\\
\a'^{3/2}\widehat{F}_3&=m_0 \tilde{\b}_1\tilde{\b}_2^{1/2}\hat{\mathfrak{e}}^1\wedge \hat{\mathfrak{e}}^2\wedge \hat{\mathfrak{e}}^3+
\tilde{\b}_1\tilde{\b}_2^{1/2}\ast_7F_4\nonumber\\
&+\a'F_2\wedge \left[v_1\tilde{\b}_1^{1/2}(\cos\psi \hat{\mathfrak{e}}^1+\sin\psi \hat{\mathfrak{e}}^2)+v_3\tilde{\b}_2^{1/2} \hat{\mathfrak{e}}^3 \right]\\
&=-\frac{\a'^3 m_0\tilde{\b}_1^2\tilde{\b}_2}{\tilde{\Delta}}v_1 dv_1\wedge dv_3\wedge\hat{\eta}+\tilde{\b}_1\tilde{\b}_2^{1/2}\ast_7F_4
+\a'^2F_2\wedge(v_1dv_1+v_3 dv_3)\nn\\
\a'^{3/2}\widehat{F}_5&=(1+\ast)\left\{\a'F_4\wedge \left[v_1\tilde{\b}_1^{1/2}(\cos\psi \hat{\mathfrak{e}}^1+\sin\psi \hat{\mathfrak{e}}^2)
+v_3\tilde{\b}_2^{1/2} \hat{\mathfrak{e}}^3 \right]+\right. \nonumber\\
&\left.\phantom{\tilde{\b}_1^{1/2}}+ \tilde{\b}_1\tilde{\b}_2^{1/2}F_2\wedge \hat{\mathfrak{e}}^1\wedge \hat{\mathfrak{e}}^2\wedge 
\hat{\mathfrak{e}}^3 \right\} \nn\\
&=-\tilde{\b}_1\tilde{\b}_2^{1/2}\ast_7 F_2+\a'^2 F_4\wedge (v_1dv_1+v_3dv_3)- \frac{v_1\a'^3 \tilde{\b}_1^2 \tilde{\b}_2}{\tilde{\Delta}}  
F_2\wedge dv_1\wedge dv_3\wedge \hat{\eta} \nonumber\\
&-\a'^3\frac{v_1\tilde{\b}_1 \tilde{\b}_2^{1/2}}{\tilde{\Delta}}(\ast_7F_4)\wedge (v_3\tilde{\b}_2dv_1- v_1\tilde{\b}_1dv_3)\wedge \hat{\eta}
\end{align}
or 
\begin{align}
\a'^{3/2}\widehat{F}_1&=\a'^2 m_0\, \rho d\rho\\
\a'^{3/2}\widehat{F}_3&=\frac{m_0 \a'^3}{\Delta}\tilde{\b}_1^2\tilde{\b}_2\r^2 d\r\wedge vol_{\tilde{S}^2}+
\tilde{\b}_1\tilde{\b}_2^{1/2}\ast_7F_4+\a'^2\r F_2\wedge d\r \nn\\
\a'^{3/2}\widehat{F}_5&=\a'^2\r F_4\wedge d\r+\frac{ \a'^3}{\Delta}\tilde{\b}_1^2\tilde{\b}_2\r^2 F_2\wedge d\r\wedge vol_{\tilde{S}^2}
-\tilde{\b}_1\tilde{\b}_2^{1/2}\ast_7F_2\nn\\
&+\frac{\a'^3}{\Delta} \tilde{\b}_1\tilde{\b}_2^{1/2}\ast_7F_4\wedge\left[-\r^2\sin^2\chi\cos\chi \left(\tilde{\b}_1-\tilde{\b}_2\right)\hat{\eta}\wedge d\r\right.\nn\\
&+\left.\r^3\sin\chi \left(\tilde{\b}_2\cos^2\chi+\tilde{\b}_1\sin^2\chi\right)\hat{\eta}\wedge d\chi
 \right]
\end{align}
\ese
where $vol_{\tilde{S}^2}=\sin\chi d\chi\wedge d\psi$ and
\begin{equation}
\hat{\mathfrak{e}}_a=e^{C_a}\tilde{\Delta}^{-1}\left[z_a z_b Dz_c+e^{2\sum_{b\neq a}C_b}Dz_a+\epsilon_{abc}z_b e^{2C_b}Dz_c \right]\; .
\end{equation}


\section{Lifshitz Solutions} \label{lifs-sol}

In \cite{Balasubramanian:2010uk, Donos:2010tu}, an infinite class of Lifshitz solutions of $D=10$ and $D=11$ 
supergravity with dynamical exponent $z=2$ was considered. In this section we review some aspects of this class of solutions  in \cite{Donos:2010tu}, 
which has a special limit the solutions of \cite{Balasubramanian:2010uk}. 

This type IIB supergravity solution has a metric of the form
\bse
\begin{equation}
 ds^2=ds_{Lif}^2+ L^2 ds_{E_5}^2
\end{equation}
where
\begin{equation}
\begin{split}
ds_{Lif}^2&=L^2\left(r^2(+2d\sigma dt+dx_1^2+dx_2^2)+\frac{1}{r^2}dr^2+f d\sigma^2\right)\\
&=L^2\left(-\frac{r^4}{f}dt^2+r^2(dx_1^2+dx_2^2)+\frac{1}{r^2}dr^2+f \left( d\sigma+\frac{r^2}{f}dt\right)^2\right),
\end{split}
\end{equation} 
where $f$ is a function of $\s$ and of the coordinates of the Sasaki-Einstein manifold $E_5$.
This background  has also the nontrivial fields\footnote{For $\tau=C_0+ie^{-\phi}$, $P=(i/2)e^\phi d\tau$ and $G=i e^{\phi/2} d\tau(\tau dB-dC_2)$, 
where the scalar $C_0$ is the axion and $\phi$ the dilaton, and the $2$-forms in the NS-NS and the RR sectors are $B$ and $C_2$, respectively.}
\begin{align}
F_5&=4L^4(1+\ast)Vol_{E_5}\; ,\\
G&= d\s\wedge W\; , \\
P&=g d\s\; .
\end{align}
\ese
We can recover the standard solution $AdS_5\times E^5$ solution by $W=f=g=0$ and the solution of \cite{Balasubramanian:2010uk} 
can be obtained in the special limit $W=0$,  $f=f(\s)>0$, $g=g(\s)\in \mathbb{R}$. In addition, the coordinate $\s$ is compact 
and parametrizes a circle $S^1$.

Another interesting class of solutions are those with constant $f$. When we set $f$ to a constant, $f=1$, the four dimensional non-compact 
part of this metric is precisely the metric with the Lifshitz symmetry for $z=2$ and $r=\frac{1}{u}$, that is
\be 
ds^2=L^2\left(-\frac{dt^2}{u^{4}}+\frac{dx^i dx^i}{u^2}+\frac{du^2}{u^2} \right).
\ee
Also, in \cite{Donos:2010tu}, the authors showed that under certain conditions, we can consider the KK-reduction 
on $S^1\times E_5$ and we get contact with the bottom-up construction of Lifshitz solutions.

\subsection{Homogeneous Space $T^{(1,1)}$}

We start considering the particular solution in which $E_5$ is the homogeneous space $(SU(2)\times SU(2))/U(1)$, that is, $E_5=T^{(1,1)}$ with metric
\begin{align}
ds_{T^{(1,1)}}^2&=\frac{1}{9}(d\psi+\cos\theta_1 d\phi_1+\cos\theta_2 d \phi_2)^2+\frac{1}{6}(d\theta_1^2+\sin^2\theta_1d\phi_1^2)
+\frac{1}{6}(d\theta_2^2+\sin^2\theta_2d\phi_2^2) \nonumber\\
&=\frac{1}{L^2}\left((\mathfrak{e}^1)^2+(\mathfrak{e}^2)^2+(\mathfrak{e}^3)^2+(\mathfrak{e}^{\hat{1}})^2+
(\mathfrak{e}^{\hat{2}})^2\right) \label{intspace}
\end{align}
where 
\begin{equation}
\begin{split}
\mathfrak{e}^1&=\frac{L}{\sqrt{6}}\tau^1\; ,\quad \mathfrak{e}^2=\frac{L}{\sqrt{6}}\tau^2 \; ,\quad \mathfrak{e}^3
=\frac{L}{3}\left(\tau^3+\cos\theta_1 d\phi_1\right)\\
\mathfrak{e}^{\hat{1}}&=\frac{L}{\sqrt{6}}d \theta_1\; ,\quad \mathfrak{e}^{\hat{2}}=\frac{L}{\sqrt{6}}\sin \theta_1 d\phi_1
\end{split}
\end{equation}
and $\tau_i$ are given by (\ref{taui})
Using these results and the notation of \cite{Kelekci:2014ima}, we see that 
\begin{equation}
\begin{split}
e^{C_1}&=e^{C_2}=\frac{L}{\sqrt{6}} \equiv \bar{\b}_1^{1/2}\; , \quad e^{C_3}=\frac{L}{3}\equiv \bar{\b}_2^{1/2}\; ,\\
\quad A^1&=A^2=0\; , \quad A^3=\cos \theta_1 d\phi_1\; .
\end{split}
\end{equation}
In the NS-NS sector we have  the Kalb-Ramond field $B$ with field strength
\begin{equation}
H_3= -\sqrt{2}k \,d\sigma \wedge\left( \mathfrak{e}^{\hat{1}}\wedge \mathfrak{e}^{\hat{2}} 
+\mathfrak{e}^{1}\wedge \mathfrak{e}^{2}\right)\; ,
\end{equation}
and this field strength can be generated by 
\begin{equation}
B_2=-\frac{k L^2}{3\sqrt{2}}\cos \theta_1\,d\sigma \wedge d\phi_1+\frac{kL^2}{3\sqrt{2}}\, d\sigma\wedge \tau_3\; ,
\end{equation}
which means that $\b_3=\frac{k L^2}{3\sqrt{2}}d\sigma$.
The dilaton and the axion are taken to be trivial. 

In the R-R sector we just have the self-dual 5-form 
\begin{equation}
\begin{split}
F_5&=4\; L^4 \left(r^3\; d\sigma\wedge dt\wedge dr\wedge dx_1\wedge dx_2+Vol_{T^{(1,1)}}\right)\\
&=4\ L^4 (1+\ast)\ Vol_{T^{(1,1)}}\; .
\end{split}
\end{equation}

Note that we can consider an ordinary T-duality and an uplift of this solution in order to find type IIA and $D=11$ solutions 
(in this particular case we have $f=k$).

\subsubsection{Nonabelian T-dual}

The nonabelian T-duality with respect to the $SU(2)$ isometry parametrized by the $(\psi, \phi_2, \theta_2)$ coordinates in 
the space $T^{(1,1)}$ has been considered in \cite{Itsios:2012zv} and was reviewed in \cite{Itsios:2013wd}. Here we consider a 
slight modification of \cite{Itsios:2012zv}, namely  now we have a non-vanishing Kalb-Ramond field and obviously the non-compact space 
is not AdS${}_5$. Then the T-dual space has metric
\begin{align}
d\hat{s}^2&=\underbrace{ds_{Lif}^2+\frac{L^2}{6}ds_1^2}_{ds_7^2}+\frac{1}{\bar\Delta}
\left[\left(\sum_{i=1}^3 z_i Dz_i\right)^2+ e^{2(C_1+C_2+C_3)}\sum_{i=1}^3e^{-2C_i}(Dz_i)^2 \right]\nn\\
&= ds_7^2+\frac{1}{\bar{\Delta}}\left\{\frac{\sqrt{2}L^2 k\a'}{3}\left(\a'^2 v_1 v_2 dv_1+(\a'^2v_3^2+\bar{\b}_1^2)dv_3\right)d\s
+\frac{k^2L^4}{18}(\a'^2v_3^2+\bar{\b}_1^2)d\s^2 \right\}\nn\\
&+\frac{\a'^2 v_1^2 \bar{\beta}_1\bar{\beta}_2}{\bar{\Delta}} \hat{\eta}^2+\frac{\a'^2}{\bar{\Delta}}\left\{(\bar{\beta}_1\bar{\beta}_2
+\a'^2 v_1^2)dv_1^2+(\bar{\beta}_1^2+\a'^2v_3^2)dv_3^2+2\a'^2 v_1v_3dv_1dv_3\right\}
\end{align}
where the coordinates are $\{z_1=\a'\hat{v}_1, z_2=\a'\hat{v}_2, z_3=\a'\hat{v}_3\}$ and their ``covariant'' derivatives $Dz_i$ are
\begin{equation}
\frac{1}{\a^\prime}Dz_1= d\hat{v}_1-\hat{v}_2 A^3\; , \quad \frac{1}{\a^\prime}Dz_2= d\hat{v}_2+\hat{v}_1 A^3\; ,\quad 
\frac{1}{\a^\prime}Dz_3= d\hat{v}_3+\frac{1}{\a'}\b_3\; ,
\end{equation}
and finally $\bar{\Delta}=[(\bar{\beta}_1^2+\a'^2v_3^2)\bar{\beta}_2+\a'^2v_1^2\bar{\beta}_1]$, which is 
related to the dual dilaton $\hat{\phi}$ field by $\bar{\Delta}=\a'^3e^{-2\hat{\phi}}$. 

The T-dual NS--NS two-form is
\begin{align}
\widehat{B}&=-\frac{k L^2}{3\sqrt{2}}\cos \theta_1\,d\sigma \wedge d\phi_1-\frac{1}{\bar{\Delta}}\left(
\epsilon_{ijk}e^{2C_i}z_i Dz_j\wedge Dz_k\right)-\cos\theta_1 Dz_3\wedge d\phi_1\;\nn \\
&=-\frac{k L^2}{3\sqrt{2}}\cos \theta_1\,d\sigma \wedge d\phi_1
-\frac{\a'^3}{\bar{\Delta}}\left( \bar{\beta}_2v_1 v_3 dv_1-\bar{\b}_1 v_1^2 dv_3\right)\wedge \hat{\eta}\; \nn\\
&-\frac{kL^2}{3\sqrt{2}\bar{\Delta}}\bar{\beta}_2(\bar{\beta}_1^2+\a'^2 v_3^2)d\s\wedge \hat{\eta}
+\frac{kL^2}{3\sqrt{2}}d\s\wedge d\psi-\a'\cos\theta_1 dv_3\wedge d\phi_1\; .
\end{align}

Using the fact that the original self-dual five-form is given by
\begin{equation}
\begin{split}
F_5&= \frac{2 L^2}{9}(1+\ast)\ \mathfrak{e}^{\hat{1}}\wedge \mathfrak{e}^{\hat{2}}\wedge \tau_1\wedge \tau_2\wedge (\tau_3+\cos \theta_1 d\phi_1)\\
&\equiv\ (1+\ast)\ G_2\wedge \tau_1\wedge \tau_2\wedge (\tau_3+\cos \theta_1 d\phi_1)\, ,
\end{split}
\end{equation}
the dual R--R sector is defined by
\bse
\begin{align} 
\a'^{3/2}\widehat{F}_2&=-G_2\\
\a'^{3/2}\widehat{F}_4&=G_2\wedge \left\{\frac{\a'v_1}{(\bar{\b}_1\bar{\b}_2)^{1/2}}(\cos\psi \hat{\mathfrak{e}}_2-\sin\psi 
\hat{\mathfrak{e}}_1)\wedge \hat{\mathfrak{e}}_3+\frac{\a' v_3}{\bar{\b}_1}\hat{\mathfrak{e}}_1\wedge \hat{\mathfrak{e}}_2 \right\} \nn\\
&=\frac{\a'^2}{\bar{\Delta}}G_2\wedge \left\{ \a'\bar{\b}_2 v_1 v_3 dv_1-\bar{\b}_1 v_1^2(\a'dv_3+\b_3)\right\}\wedge \hat{\eta}\;,
\end{align}
\ese
where
\begin{equation}
\hat{\mathfrak{e}}_a=e^{C_a}\bar{\Delta}^{-1}\left[z_a z_b Dz_b+e^{2\sum_{b\neq a}C_b}Dz_a+\epsilon_{abc}z_b e^{2C_b}Dz_c \right]\; ,
\end{equation}
which implies
\bse
\begin{align}
\hat{\mathfrak{e}}'_1&\equiv\cos \psi \hat{\mathfrak{e}}_1+\sin\psi \hat{\mathfrak{e}}_2\nn\\
&=\frac{\bar{\b}_1^{1/2}}{\bar{\Delta}}\left[\a'(\bar{\b}_1\bar{\b}_2+\a'^2v_1^2)dv_1+\a'^2v_1v_3 (\a'dv_3+\b_3)-\a'^2 \bar{\b}_2 v_1 v_3\hat{\eta}  \right]\\
\hat{\mathfrak{e}}'_2&\equiv-\sin \psi \hat{\mathfrak{e}}_1+\cos\psi \hat{\mathfrak{e}}_2\nn\\
&= \frac{\bar{\b}_1^{1/2}}{\bar{\Delta}}\left[\a'^2 \bar{\b}_2 v_3 dv_1-\a'\bar{\b}_1 v_1 (\a' dv_3+\b_3)+\a'\bar{\b}_1 \bar{\b}_2 v_1 \hat{\eta}  \right]\\
\hat{\mathfrak{e}}_3&=\frac{\bar{\b}_2^{1/2}}{\bar{\Delta}}\left[\a'^3 v_1 v_3 dv_1+(\bar{\b}_1^2+\a'^2v_3^2)(\a' dv_3
+\b_3)+\a'^2\bar{\b}_1v_1^2 \hat{\eta} \right]\; .
\end{align}
\ese
Here $\b_3=\frac{k L^2}{3\sqrt{2}}d\s$.

\subsection[Sasaki-Einstein Space]{Sasaki-Einstein Space $Y^{p,q}$}

Another possible choice is $E_5=Y^{p,q}$, such that the Sasaki-Einstein metric is 
\begin{subequations}
\begin{align}
ds_{Y^{p,q}}^2&=w(y)\left(d\alpha+h(y)\tau_3 \right)^2+\frac{1-y}{6}(\tau_1^2+\tau_2^2)+\frac{dy^2}{w(y) q(y)}+\frac{q(y)}{9}\tau_3^2\nonumber\\
&=g(y)\left(\tau_3+\frac{w(y)h(y)}{g(y)}d\alpha\right)^2+\frac{w(y) q(y)}{9 g(y)}d\a^2+\frac{1}{w(y) q(y)}dy^2+\frac{1-y}{6}(\tau_1^2+\tau_2^2)
\end{align}
where $g(y)=q(y)/9+w(y)h(y)^2$ and
\begin{equation}
h(y)=\frac{a-2y+y^2}{6(a-y^2)}\; , \quad w(y)=\frac{2(a-y^2)}{1-y}\; ,\quad q(y)=\frac{a-3y^2+2y^3}{a-y^2}\; .
\end{equation}
\end{subequations}
Here $a$ is a real constant. As studied in \cite{Gauntlett:2004yd}, in these manifolds there is a $2$-sphere fibration parametrized 
by $(y, \psi)$, with $y\in [y_1,y_2]$ over a $2$-sphere parametrized by $(\theta, \phi)$. Also, the coordinate $\a$ parametrizes a 
circle of length $2\pi l_\a$. In these spaces, we have
\begin{align}
a&=\frac{1}{2}-\frac{p^2-3q^2}{4 p^3}\sqrt{4p^2-3q^2}\nonumber\\
y_1&=\frac{1}{4p}\left(2p-3q-\sqrt{4p^2-3q^2} \right)\\
y_2&=\frac{1}{4p}\left(2p+3q-\sqrt{4p^2-3q^2} \right)\nonumber\; ,
\end{align}
where $(p,q)$ are relative integers and $p>q>0$.

We can set the axion and the dilaton to be zero, but, contrary to the previous solution, this condition does not imply 
that the function $f$ is a constant. In fact, it satisfies the following equation
\be 
-4 f+\frac{2}{1-y}\d_y[(a-3y^2+2y^3)\d_y f]+\frac{1}{(1-y)^4}=0.
\ee

We define the vielbeins
\begin{subequations}
\begin{align}
\mathfrak{e}^\a&=\frac{L}{3}\sqrt{\frac{wq}{g}}d\a\; ,\quad \mathfrak{e}^y=\frac{L}{\sqrt{wq}}dy \\
\mathfrak{e}^1&=L\sqrt{\frac{(1-y)}{6}}\tau_1\; ,\quad \mathfrak{e}^2=L\sqrt{\frac{(1-y)}{6}}\tau_2\; , \quad
\mathfrak{e}^3=L\sqrt{g}\left( \tau_3+\frac{wh}{g} d\a\right)\; .
\end{align}
\end{subequations}
In this $Y^{p,q}$ background, one has
\begin{equation}
\begin{split}
W&=-\frac{L^2}{\sqrt{72}}d\left[\frac{1}{1-y}(6d\alpha+\tau_3)\right]=
-\frac{L^2}{6\sqrt{2}(1-y)}\left[\frac{1}{(1-y)}dy\wedge(6d\alpha+\tau_3)+\tau_1\wedge \tau_2\right]\\
&=-\frac{L^2}{6\sqrt{2}(1-y)^2}\left\{dy\wedge \left[\left(6-\frac{wh}{g}\right)d\alpha+\frac{1}{L\sqrt{g}}\mathfrak{e}^3\right]
+\frac{6}{L^2}\; \mathfrak{e}^1\wedge \mathfrak{e}^2\right\}\;,
\end{split}
\end{equation}
which means that
\begin{align}
H&=-d\sigma\wedge W\nonumber\\
&=\frac{L^2}{6\sqrt{2}(1-y)^2}\left(6-\frac{wh}{g}\right)d\sigma\wedge dy \wedge d\a+\frac{L}{6\sqrt{2g}(1-y)^2}d\s \wedge dy \wedge \mathfrak{e}^3\\
&+\frac{1}{\sqrt{2}(1-y)^2}d\s \wedge \mathfrak{e}^1\wedge \mathfrak{e}^2\; .\nonumber
\end{align}
Using the notation of \cite{Kelekci:2014ima}, we find
\begin{subequations}
\begin{align}
e^{C_1}&=e^{C_2}=L\sqrt{\frac{1-y}{6}}=\breve{\b}_1^{1/2}\; , \quad e^{C_3}=L\sqrt{g}=\breve{\b}_2^{1/2}\; , \\
A_1&=A_2=0\; , \quad A_3=\frac{wh}{g}d\a\; , \\
\b_1&=\b_2=0\; , \quad \b_3=-\frac{L^2}{6\sqrt{2}(1-y)}d\s\; ,\quad D \b_3=d \b_3=-\frac{L^2}{6\sqrt{2}(1-y)^2}dy\wedge d\s\cr
b_i&=0\; ,
\end{align}
\end{subequations}
therefore
\begin{equation}
B=-\frac{L^2}{\sqrt{2}(1-y)}d\sigma\wedge d\a- \frac{L^2}{6\sqrt{2}(1-y)}d\sigma\wedge \tau_3\; .
\end{equation}
The R--R five-form for the solution is
\begin{align}
F_5&=4L^4\ (1+\ast)\ Vol_{Y^{p,q}}\\
&=\frac{2L^2}{3}(1-y)\sqrt{g}\ (1+\ast)\ \mathfrak{e}^\a\wedge \mathfrak{e}^y\wedge \tau^1 \wedge \tau^2\wedge \left(\tau^3+\frac{wh}{g}d\a\right)\\
&= (1+\ast)\ \breve{G}_2\wedge \tau_1\wedge \tau_2\wedge \left(\tau^3+\frac{wh}{g}d\a\right)\; .
\end{align}
Finally, the metric of the T-dual space becomes
\begin{align}
d\hat{s}^2&=\underbrace{ds_{Lif}^2+\frac{wqL^2}{9g}d\a^2+\frac{L^2}{wq}dy^2}_{d\breve{s}_7^2}
+\frac{1}{\breve{\Delta}}\left[\left(\sum_{i=1}^3 z_i Dz_i\right)^2+ e^{2(C_1+C_2+C_3)}\sum_{i=1}^3e^{-2C_i}(Dz_i)^2 \right]\nn\\
&= d\breve{s}_7^2+\frac{1}{\breve{\Delta}}\left\{-\frac{\a'}{3\sqrt{2}(y-1)}\left(\a'^2 v_1 v_3 dv_1+(\a'^2v_3^2
+\breve{\b}_1^2)dv_3\right)d\s\right.\cr
&\left.+\frac{(\a'^2v_3^2+\breve{\b}_1^2)}{72(1-y)^2}L^4d\s^2 \right\}
+\frac{\a'^2 v_1^2 \breve{\beta}_1\breve{\beta}_2}{\breve{\Delta}} \hat{\g}^2\cr
&+\frac{\a'^2}{\breve{\Delta}}
\left\{(\breve{\beta}_1\breve{\beta}_2+\a'^2 v_1^2)dv_1^2+(\breve{\beta}_1^2+\a'^2v_3^2)dv_3^2+2\a'^2 v_1v_3dv_1dv_3\right\}\; ,
\end{align}
where $\hat{\g}=d\psi+\frac{hw}{g}d\a$ and $\breve{\Delta}=(\breve{\b}_1^2+\a'^2v_3^2)\breve{\b}_2+\a'^2 v_1^2\breve{\b}_1$. The NS--NS two-form is
\begin{align}
\widehat{B}&=-\frac{L^2}{\sqrt{2}(1-y)}d\sigma\wedge d\a-\frac{1}{\breve{\Delta}}\left(
\epsilon_{ijk}e^{2C_i}z_i Dz_j\wedge Dz_k\right)- Dz_3\wedge A_3\;\nn \\
&=\frac{L^2}{\sqrt{2}(1-y)}\left(1-\frac{hw}{6g}\right)d\a\wedge d\s-\frac{\a'^2}{\breve{\Delta}}\left[\a' \breve{\beta}_2 v_1 v_3 dv_1
-\breve{\beta}_1 v_1^2 (\a' dv_3+\breve{\b}_3)\right]\wedge \hat{\g}\; \nn\\
&+\frac{\a' h w}{g}d\a\wedge dv_3\; ,
\end{align}
and the dilaton is
\be
\hat{\phi}=-\frac{1}{2}\ln \left(\frac{\breve{\Delta}}{\a'^3} \right)\; .
\ee
The fields of the T-dual R--R sector are
\bse
\begin{align} 
\a'^{3/2}\widehat{F}_2&=-\breve{G}_2\\
\a'^{3/2}\widehat{F}_4&=\breve{G}_2\wedge \left\{\frac{\a' v_1}{(\breve{\b}_1\breve{\b}_2)^{1/2}}(\cos\psi \hat{\mathfrak{e}}_2
-\sin\psi \hat{\mathfrak{e}}_1)\wedge \hat{\mathfrak{e}}_3+\frac{\a'v_3}{\breve{\b}_1}\hat{\mathfrak{e}}_1\wedge \hat{\mathfrak{e}}_2 \right\}\nn\\
&=\frac{\a'^2}{\breve{\Delta}}\breve{G}_2\wedge \left\{ \a'\breve{\b}_2 v_1 v_3 dv_1-\breve{\b}_1 v_1^2(\a'dv_3+\breve{\b}_3)\right\}\wedge \hat{\g}\;,
\end{align}
\ese
with the vielbeins
\bse
\begin{align}
\hat{\mathfrak{e}}'_1&\equiv\cos \psi \hat{\mathfrak{e}}_1+\sin\psi \hat{\mathfrak{e}}_2\nn\\
&=\frac{\breve{\b}_1^{1/2}}{\breve{\Delta}}\left[\a'(\breve{\b}_1\breve{\b}_2+\a'^2v_1^2)dv_1+\a'^2v_1v_3 (\a'dv_3+\breve{\b}_3)
-\a'^2 \breve{\b}_2 v_1 v_3\hat{\g}  \right]\\
\hat{\mathfrak{e}}'_2&\equiv-\sin \psi \hat{\mathfrak{e}}_1+\cos\psi \hat{\mathfrak{e}}_2\nn\\
&= \frac{\breve{\b}_1^{1/2}}{\breve{\Delta}}\left[\a'^2 \breve{\b}_2 v_3 dv_1-\a'\breve{\b}_1 v_1 (\a' dv_3+\breve{\b}_3)
+\a'\breve{\b}_1 \breve{\b}_2 v_1 \hat{\g}  \right]\\
\hat{\mathfrak{e}}_3&=\frac{\breve{\b}_2^{1/2}}{\breve{\Delta}}\left[\a'^3 v_1 v_3 dv_1+(\breve{\b}_1^2+\a'^2v_3^2)(\a' dv_3+\breve{\b}_3)
+\a'^2\breve{\b}_1v_1^2 \hat{\g} \right]\; .
\end{align}
\ese
Here $\breve{\b}_3=\frac{ -L^2}{6\sqrt{2}(1-y)}d\s$.


\section{Holographic Dual Field Theory}

\subsection{Quantized Charges for Galilean Solutions}

In \cite{Lozano:2014ata}, the authors considered the dualization of the background 
holographic dual to the ABJM theory \cite{Aharony:2008ug}, which consists 
of a metric for $AdS_4\times \mathbb{CP}^3$ in type IIA, together with two R--R fields, $F_2$ and $F_4$. They also 
calculated the conserved charges of the dual background.

Considering the effect of the non-relativistic deformation of the ABJM background considered in \cite{Singh:2010rt}, we compute the 
conserved charges of the background that we found in the last section. We compare our results with \cite{Lozano:2014ata} in order 
to see the effect of the non-relativistic deformation of the background \cite{Aharony:2008ug}.
We calculate the conserved charges of the solutions in sections 2.1 (massless type IIA ) and 2.2 (massive type IIA) separately.

\subsubsection{Massless type IIA}

We start with a short review of the conserved charges of $AdS_4\times \mathbb{CP}^3$ and its NATD solution. 
The solution has the metric of $AdS_4\times \mathbb{CP}^3$, the dilaton 
$\phi=\ln \left(R/k\right)$ and the R--R forms \cite{Lozano:2014ata}
\bse
\begin{align}
dC_{(1)}&=2kd\omega\\
dC_{(3)}&=\frac{3}{8}k L^2 Vol_{AdS_4}
\end{align}
\ese
in such a way that 
\be 
\int_{\mathbb{CP}^1}dC_{(1)}=2\pi k\; \quad \text{and}\quad \frac{1}{(2\pi \a'^{1/2})^5}\int_{\mathbb{CP}^3}\ast 
\mathcal{F}_4=N\in \mathbb{Z}\; \Rightarrow \; L^4=\frac{32\pi^2 \a'^{5/2}}{k}N\; ,
\ee
where we have used the $\mathbb{CP}^1$ defined by $(\xi=\pi/2, \theta_2, \phi_2)$, and
$\int Vol_{\mathbb{CP}^3}=\frac{\pi^3}{6}$. We see that these quantization
conditions agree perfectly with the quantization conditions (\ref{quant.01}-\ref{quant.02}) of the Galilean solution.

In \cite{Lozano:2014ata}, the authors calculated the charge $N_{D5}$ in the dual field theory, which was found from 
an integration of the dual $3$-form over the cycle defined by 
$\Sigma^3=(\zeta, \theta_1, \phi_1)$, such that\footnote{In their notation $\a'=1$.}
\be 
N_{5}=\frac{kL^4}{64\pi \a'^{5/2}}.
\ee
In our case of the Galilean solution of massless type IIA, 
we must consider the same calculation for the dual $\widehat{\mathcal{F}}_3$. Using that
\be 
\left. \phantom{\frac{1}{1}} \widehat{\mathcal{F}}_3\right|_{\Sigma^3}= \frac{6\b_1\b_2^{1/2}k}{R^2\a'^{3/2}}
\mathfrak{e}^\zeta \wedge \mathfrak{e}^\theta \wedge \mathfrak{e}^\phi=\frac{3kR^4}{16\a'^{3/2}}\sin^3\zeta \cos^3\zeta d\zeta\wedge vol_{S_1^2}  \; ,
\ee
where $vol_{S_1^2}=\sin\theta_1 d\theta_1\wedge d\phi_1$,  we compute the charge
\be
\int_{\Sigma^3}\widehat{\mathcal{F}}_3=\frac{k \pi R^4}{16\a'^{3/2}}.
\ee
Imposing the quantization condition for the Page charge, 
\be
\frac{1}{(2\pi \a'^{1/2})^2}\int_{\Sigma_3}{\cal F}_3=\hat Q_{D5}\in \mathbb{Z}\;,
\ee
we obtain
\be
\widehat{\mathcal{Q}}_{D5}=\frac{k R^4}{64\pi \a'^{5/2}} .
\ee

But since originally $R^4$ satisfied the relation $k R^4=32 \pi^2 \a'^{1/2}N$, the charge $\widehat{\mathcal{Q}}_{D5}$ cannot be an integer, 
and in this case, the radius $R$ in the dual theory will be defined through new relations. The non-integer charge in the non-abelian 
T-dual theory is a generic feature which arises from the violation of the condition $T_{D(p-n)}=(2\pi)^nT_{Dp}$.

In the present case, we see that if we consider the $5$-cycle $\Sigma^5=(\zeta, \theta_1, \phi_1, \chi, \psi)\equiv
 (\zeta, \theta_1, \phi_1, v_1=n\pi \sin \xi, v_3=n \pi \cos \xi, \psi)$ in the T-dual background, we compute the restriction
\begin{align} 
\left.\a'^{3/2}\mathcal{\widehat{F}}_5  \phantom{\frac{1}{1}}\right|_{\Sigma^5} &=n \pi \a' \mathcal{\widehat{F}}_3\wedge vol_{\tilde{S}^2}\; ,
\end{align}
which is consistent with a large gauge transformation $\mathcal{\widehat{F}}_5\; \to\; \mathcal{\widehat{F}}_5
+n\pi\a' \mathcal{\widehat{F}}_3\wedge vol_{\tilde{S}^2}$ (with the volume form $vol_{\tilde{S}^2}=\sin\chi d\chi\wedge d\psi$), and therefore  we find
\be
\mathcal{Q}_{D3}=n \mathcal{Q}_{D5}\; .
\ee

The field theory on the boundary is a 2+1 dimensional CS gauge theory, as was the ABJM theory before the NATD. The CS gauge groups have levels, that 
should be possible to calculate from the gravity dual.
As in \cite{Lozano:2014ata}, we can define the levels of the AdS/CFT dual field theory as
\be 
q_5=\left|\frac{1}{(2\pi\a'^{1/2})^2}\int_{\tilde{\Sigma}^3}\widehat{\mathcal{F}}_3\right|\; , \qquad q_3
=\left|\frac{1}{(2\pi\a'^{1/2})^4}\int_{\tilde{\Sigma}^5}\widehat{\mathcal{F}}_5\right|\;,
\ee
where the integrations are performed on the cycles $\tilde{\Sigma}_3=(\rho, \theta_1, \phi_1)\equiv (v_3, \theta_1, \phi_1)$ 
and $\tilde{\Sigma}_5=(\rho, \theta_1, \phi_1, \chi, \psi)\equiv (\theta_1, \phi_1, v_1, v_3 ,\psi)$, respectively. 
In the presence of a large gauge transformation, one obtained in the case in \cite{Lozano:2014ata}
\be 
q_5=k\frac{(2n+1)\pi}{4\a'^{1/2}}\; , \qquad q_3=k\frac{(3n+2)\pi}{12\a'^{1/2}}\; .
\ee
Using the same definitions, in our case we obtain from (\ref{hatfthree}) 
\bse
\be 
\left.\phantom{\frac{1}{1}}\a'^{3/2}\widehat{F}_3\right|_{\tilde{\Sigma}_3}=-\frac{\a'^2 k}{2}\rho\, vol_{S_1^2}\wedge  d\rho \phantom{\frac{1}{1}} \; ,  \;
\left.\phantom{\frac{1}{1}}\a'^{3/2}\widehat{F}_1\wedge \widehat{B}\right|_{\tilde{\Sigma}_3}=0\; ,
\ee
therefore we obtain in a similar manner to the above case
\be 
 k_5=k\frac{(2n+1)\pi}{4\a'^{1/2}}\; .
\ee
\ese
We also find from (\ref{hatfthree}) that
\bse
\be 
\left.\a'^{3/2}\widehat{\cal F}_5\right|_{\tilde{\Sigma}_5}=\frac{k\a'^3}{2}\r^2 d\r\wedge vol_{S_1^2}\wedge vol_{\tilde{S}^2}\; ,
\ee
and given that
\be 
\left.\a'^{3/2} F_3\wedge vol_{\tilde{S}^2}\right|_{\tilde{\Sigma}_5}=-\frac{k\a'^2}{2}\r d\r \wedge vol_{S_1^2}\wedge vol_{\tilde{S}^2}\; 
\ee
under a large gauge transformation $\widehat{\mathcal{F}}_5\to\widehat{\mathcal{F}}_5+n\pi\a' \widehat{F}_3\wedge vol_{\tilde{S}^2}$, we have
\be 
\int_{\tilde{\Sigma}^5}(\widehat{\mathcal{F}}_5+n\pi\a' \widehat{F}_3\wedge vol_{\tilde{S}^2})=k\frac{(4\pi)^2}{12\a'^{1/2}}(3n+2)\pi^3\; .
\ee
\ese
Then finally
\be 
 k_3=k\frac{(3n+2)\pi}{12\a'^{1/2}}\; ,
\ee
such that $(3n+2)k_5=3(2n+1)k_3$. Using these relations, we find the relations between the radius $R$ and the quantized charges of the background
\bse
\begin{align}
	R^4 k_5&=32 \pi^2\a'^2\left(\widehat{\cal Q}_{D3}+\frac{1}{2}\widehat{\cal Q}_{D5}\right)\\
	R^4 k_3&=16 \pi^2\a'^2\left(\widehat{\cal Q}_{D3}+\frac{2}{3}\widehat{\cal Q}_{D5}\right).
\end{align}
\ese
If we compare our results with \cite{Lozano:2014ata} we can see that the non-relativistic deformation does 
not change the quantization condition of the theory and of its non-abelian T-dual.

\subsubsection{Massive type IIA}

We now turn to the model in section 2.2.
We first consider the model before the T-duality. We find
\bse
\be 
\ast_{10} F_4=\frac{5^3\sqrt{5}m_0}{f^6}vol_{\mathbb{CP}^3}\;,
\ee
giving
\be 
{\cal Q}_{D2}^{m_0}=\frac{5^3\sqrt{5}}{192\pi^2\a'^{5/2}}\frac{m_0}{f^6}\; .
\ee
\ese
In \cite{Singh:2010rt}, the author considered the compactification of the ordinary type IIA theory and of the massive type IIA theory 
to four dimensions.\footnote{Since the only relationship between these two theories is Hull's duality \cite{Hull:1998vy}, the author argued 
that the similarity between the 4D actions means that there is a mapping between the Romans' mass and the flux $k$. 
In that case, $f\propto 1/R$ and $m_0\propto k/R^2$, so that, up to numerical constants, one could write ${\cal Q}_{D2}^{m_0}\propto {\cal Q}_{D2}$.}

We calculate the D$5$-charge by using (\ref{ffour}) to write 
\be 
\left.\left.\phantom{\frac{1}{1}}\widehat{{\cal F}}_3^{m_0}\right|_{\Sigma^3}=\frac{1}{\a'^{3/2}}\tilde{\b}_1\tilde{\b}_2^{1/2}\ast_7 F_4\right|_{\Sigma^3}
=\frac{\sqrt{5}m_0}{\a'^{3/2}}\tilde{\b}_1\tilde{\b}_2^{1/2}\mathfrak{e}_{m_0}^\zeta\wedge\mathfrak{e}_{m_0}^\theta\wedge
\mathfrak{e}_{m_0}^\phi\; ,
\ee
where $*_7$ is Poincar\'{e} duality in $ds_7^2$. We then calculate the magnetic D5-charge associated with this flux as
\be 
\widehat{{\cal Q}}_{D5}^{m_0}=\frac{5^3\, \sqrt{5}}{384\pi \a'^{5/2}}\frac{m_0}{f^6}\; .
\ee
For the cycle $\Sigma^5$, we have $\widehat{{\cal F}}_5^{m_0}|_{\Sigma^5}=0$, so now we obtain $\widehat{{\cal Q}}_{D3}^{m_0}
=n\widehat{{\cal Q}}_{D5}^{m_0}$ when we consider a large gauge transformation.

On the other hand, $\widehat{{\cal F}}_3^{m_0}|_{\tilde{\Sigma}^3}=0$, which remains equal to zero after a large gauge transformation. 

We can define a third cycle $\check{\Sigma}^3=(\r, \chi,\psi)$, which gives
\be 
\left.\phantom{\frac{1}{1}}\widehat{{\cal F}}_3^{m_0}\right|_{\check{\Sigma}^3}=0,
\ee
but after a large gauge transformation $\widehat{{\cal F}}_3^{m_0}\to \widehat{{\cal F}}_3^{m_0}+n\pi\a'\widehat{{\cal F}}_1^{m_0}\wedge 
vol_{\tilde{S}^2}$, we find
\be 
k_5^{m_0}=\frac{m_0\pi^2 \a'^{1/2}}{2}(2n+1)\; .
\ee
Finally, for the cycle $\tilde{\Sigma}^5=(\theta_1, \phi_1, \r , \chi, \psi)$, we have that $k_3=0$ even after a large gauge transformation.

\subsection{Quantized Charges for Lifshitz Solutions}

\subsubsection[Homogeneous Space SU(2)XSU(2)/U(1)]{Homogeneous Space $SU(2)\times SU(2)/U(1)$}

For the background (\ref{intspace}) in section 3.1 we start with a $5$-form
\be
F_5=\frac{2 L^4}{9}(1+\ast)vol_{S_1^2}\; \wedge \tau_1\wedge \tau_2 \wedge (\tau_3+\cos\theta_1 d\phi_1)\;,
\ee
and using similar methods we find the quantized charge 
\be 
N_{D3}=\frac{4}{27\pi }\frac{L^4}{\a'^2}.
\ee
After the T-duality we obtain the charges
\be 
L^4=\frac{27 \a'^2}{2}\widehat{{\cal Q}}_{D6}\, ,
\ee
on the cycle $S^2=(\theta_1, \phi_1)$. Using the fact that $\widehat{\cal F}_4=0$, after a large gauge transformation 
$\widehat{\cal F}_4\to \widehat{\cal F}_4+n\pi \a' \widehat{\cal F}_2\wedge vol_{\tilde{S}_2}$, we find  
$\widehat{{\cal Q}}_{D4}=n\widehat{{\cal Q}}_{D6}\; $ on the cycle $\S^4=(\theta_1, \phi_1, \chi, \psi)$,.

\subsubsection{Sasaki-Einstein Space}

In the Sasaki-Einstein case in section 3.2, we have a similar situation. The quantized charge before the T-duality is
\be 
\breve{N}_{D3}=\frac{1}{4\pi^4}\frac{L^4}{\a'^2}V_{Y^{p,q}}\, ,
\ee
on the cycle $\S^2=(\a, y)$ and  
\be 
V_{Y^{p,q}}=\int_{Y^{p,q}} Vol_{Y^{p,q}}=\frac{8\pi^3 l_\a}{3}\int_{y_1}^{y_2}dy (1-y)\; .
\ee
Repeating the previous analysis, we find
\be 
\widehat{{\cal Q}}_{D6}=\frac{L^4}{4 \pi^3 \a'^2}V_{Y^{p,q}}
\ee
on the cycle $\Sigma^2=(\a, y)$. 

Again we can use the same arguments from the previous subsection to find $\widehat{\cal F}_4=0$ on $\S^4=(\a, y, \chi, \psi)$. 
If we take a large gauge transformation $\widehat{\cal F}_4\to \widehat{\cal F}_4+n\pi \a' \widehat{\cal F}_2\wedge vol_{\tilde{S}_2}$, 
where $\tilde{S}_2=(\chi, \psi)$, we also find $\widehat{{\cal Q}}_{D4}=n\widehat{{\cal Q}}_{D6}\; $.

\subsection{Wilson Loops}

One can in principle define a Wilson loop variable in the case of nonrelativistic gravity duals, even though it is not really clear what it would mean 
in the field theory. However, we can simply calculate the observable, and leave for later issues of interpretation. 

One way to embed the Schr\"odinger algebra with $z=2$ (a particular case of conformal Galilean algebra) into string theory is to consider 
a DLCQ of a known duality \cite{Maldacena:2008wh,Herzog:2008wg,Adams:2008wt,Hartnoll:2009sz}. The general conformal Galilean 
algebra is realized holographically through the metric
\be
ds^2=L^2\left[-\frac{dt^2}{r^{2z}}+\frac{2dtd\xi+d\vec{x}^2}{r^2}+\frac{dr^2}{r^2}\right].
\ee

For the Lifshitz case, gravity duals are instead usually of the type
\be
ds^2=L^2\left[-\frac{dt^2}{u^{2z}}+\frac{d\vec{x}^2}{u^2}+\frac{du^2}{u^2}\right].
\ee
However, in \cite{Balasubramanian:2010uk,Donos:2010tu} it was suggested that for $d=4$ and $z=2$, the case considered in section 3,
we can consider the gravity dual
\begin{equation}
ds_{Lif}^2=L^2\left(r^2(-2d\sigma d\tau+dx_1^2+dx_2^2)+\frac{1}{r^2}dr^2+f d\sigma^2\right)\;,
\end{equation}
and for $\s=x^+$ and $\tau=x^-$, $\sigma$ must be a compact coordinate to obtain a 2+1 dimensional field theory dual with coordinates 
$\tau, x_1,x_2$. 

Note that compared with the Schr\"{o}dinger case, the roles of $x^+$ and $x^-$ are interchanged and $x^+$ is compact. 

\subsubsection{Wilson Loops in conformal Galilean spacetime}

The general prescription for the calculation of Wilson lines in relativistic field theories is well known \
\cite{Aharony:1999ti,smilga2001lectures,nastase2015introduction,erdmenger2015,Nastase:2007kj,Sonnenschein:1999if,Nunez:2009da,Araujo:2014kda}. 
Recently, important hints in the identification of the dual field theory of non-relativistic systems were studied in 
\cite{Danielsson:2009gi, Kluson:2009vy, Fadafan:2009an, Siahaan:2011sw, Fadafan:2015iwa}.

We want to consider the Wilson loops for the conformal Galilean gravity dual case in section 2. This formalism was also considered in \cite{Siahaan:2011sw}.

Considering a probe string which is not excited in the internal space directions, our gravity dual manifold is of the general form (without the internal space)
\be 
ds^2=\frac{R^2}{r^2}\left(-\frac{dt^2}{r^{2(z-1)}} + 2 d\xi dt+d\vec{x} \cdot d\vec{x} \right)+\frac{R^2}{r^2}dr^2\; ,
\ee
with $\xi$ compact and null, for $z=3$ (thus is not of the Schr\"{o}dinger form, which would correspond to $z=2$). 
We consider the following ansatz
\be 
t=\tau\, ,\quad x=x(\s)\, ,\quad r=r(\s)\; , \quad \xi=constant\; .
\ee
Given that the induced metric on the world-sheet is $G_{\a \b}=g_{\mu\nu}\d_\a X^\mu \d_\b X^\nu$, the Nambu-Goto action becomes
\be 
S=-\frac{1}{2\pi \a'}\int_0^T d\tau \int d\s\sqrt{-\det G}=-\frac{T R^2}{2\pi \a'} \int d\s\sqrt{\frac{(\d_\s r)^2+(\d_\s x)^2}{r^{2(z+1)}}}\; .
\ee
As usual, the analysis of the differential equations (see \cite{Sonnenschein:1999if, Nunez:2009da, Kluson:2009vy}) shows that the 
separation between the endpoints of the $\cap$--shaped string that extends from the point $x=-\ell/2$ to the point $x=\ell/2$ at the boundary $r=0$ is
\bse
\be 
\ell=2\int_0^{r_{max}} dr \frac{H(r_{max})}{\sqrt{H(r)^2-H(r_{max})^2}}\; ,
\ee
with $H^2=R^4/r^{2(z+1)}$. Therefore 
\be 
\ell(r_{max}, z)=2r_{max}\sqrt{\pi} \frac{\G \left(\frac{z+2}{2z+2} \right)}{\G\left(\frac{1}{2z+2} \right)} ,\label{distance}
\ee
\ese
and we can invert this expression, giving $r_{max}=r_{max}(\ell)$. For $z=3$, we obtain
\be
l=2r_{max}\sqrt{\pi}\frac{\Gamma\left(\frac{5}{8}\right)}{\Gamma\left(\frac{1}{8}\right)}.
\ee

The general formalism \cite{Sonnenschein:1999if, Nunez:2009da} allows us to compute a would-be quark-antiquark potential, which gives
\be 
V_{q\bar{q}}=\frac{2R^2\sqrt{\pi}}{r_{max}^z(2z+2)}\frac{\G \left(\frac{-z}{2z+2} \right)}{\G\left(\frac{1}{2z+2} \right)}\; .
\ee
It is not clear what would be the interpretation of this quantity in the field theory, since it was defined for relativistic gauge theories. 
But we can continue with the assumption that it still gives the potential between external "quarks" introduced in the theory, and see what we can deduce 
from it.

Therefore, if we consider the solutions in the section \ref{galilean}, with $z=3$, the potential is 
\be 
V_{q\bar{q}}=-\frac{2R^2\sqrt{\pi}}{3 r_{max}^3}\frac{\G \left(\frac{5}{8} \right)}{\G\left(\frac{1}{8} \right)}\; ,
\ee
which implies
\be 
\frac{dV}{d\ell}=\frac{R^2}{r_{max}^4}>0\; .
\ee
This means that the would-be quark-antiquark interaction is atractive everywhere \cite{Siahaan:2011sw, Bachas:1985xs, Arias:2009me}. 
We also have
\be 
\frac{d^2V}{d\ell^2}=-\frac{2R^2}{\sqrt{\pi} r_{max}^5} \frac{\G \left(\frac{1}{8}\right)}{\G \left(\frac{5}{8}\right)} <0  \; ,
\ee
and this condition means that the force is a monotonically non-increasing function of their separation.

\subsubsection{Wilson Loops in Lifshitz spacetime}

Consider the spacetime metric\footnote{We changed the notation $\s\to \xi$, and we keep the symbol $\s$ to the spacelike 
worldsheet coordinate. Also, we renamed $r\to \frac{1}{r}$.} of the form in the section \ref{lifs-sol},
\begin{equation}
ds_{Lif}^2=\frac{L^2}{r^2}(-2d\xi dt+dx^2+dy^2)+\frac{L^2}{r^2}dr^2+L^2 f(\xi) d\xi^2\; ,\\\label{Lif}
\end{equation}
where the coordinate $\xi$ parametrizes the circle.

First, we notice that due to the absence of the component $g_{tt}$ in the metric above, we cannot find a string configuration such that
\be 
t=\tau\; ,\quad x=x(\s)\; ,\quad r=r(\s)\; ,\quad \xi=constant\; ,
\ee
so one might consider an ansatz with the string moving also on the compact coordinate $\xi$, despite the fact that its physical meaning is rather uncertain \cite{Hartnoll:2009sz}.

We consider the following ansatz (see \cite{Kluson:2009vy}, for similar considerations in spacetimes with Schr\"odinger symmetry)
\be 
t=\tau\; ,\quad \xi=\xi(\tau) \; , \quad x=x(\s)\; ,\quad r=r(\s)\; .\label{dependence}
\ee

Then the components of the induced metric are
\be 
G_{\tau \tau}=-\frac{2L^2 }{r^2}\; \d_\tau \xi+L^2 f(\xi)(\d_\tau \xi)^2\; , \quad G_{\s\s}=\frac{L^2}{r^2}\left((x')^2+(r')^2\right)\; ,
\ee
where $x'\equiv \d_\s x$, $r' \equiv \d_\s r$ and $G\equiv \det G_{\a\b}=G_{\tau \tau}G_{\s\s}$. The Nambu-Goto action is given by
\be 
S=-\frac{1}{2\pi \a' }\int d\tau d\s \sqrt{g^2(\s,\tau)\left((x')^2+(r')^2\right)}\;,
\ee
where 
\be
g^2=-G_{\tau \tau}L^2/r^2. 
\ee

We consider the equation of motion for $\xi$,
\be 
\d_\tau\left[ G^{\tau \tau}\sqrt{-G}\left( -\frac{1}{r^2}+f(\xi)\d_{\tau}\xi \right) \right]=0\;\Rightarrow 
\d_\tau\left( \frac{G^{\tau \tau}\sqrt{-G} }{r^2}\right)=\d_\tau \left( G^{\tau \tau}\sqrt{-G} \ f (\xi)\d_{\tau}\xi\right)\; ,\label{field-eq-xi}
\ee
where $G^{\tau \tau}\equiv G_{\tau \tau}^{-1}$. From (\ref{dependence}) we see that $r$ is independent of $\tau$ and from 
\cite{Donos:2010tu}, we already know that the function $f$ does not have functional dependence on $r$. Therefore, both sides 
in (\ref{field-eq-xi}) must vanish independently. The left-hand side of the equation (\ref{field-eq-xi}), implies that 
$\frac{G^{\tau \tau}\sqrt{-G}}{r^2}=h_0(\s)$\;  therefore we take $\xi=v_\xi \tau$, where $v_{\xi}$ is a constant. 
The right-hand side gives $\d_\tau f(\xi)=0$ and since $f$ cannot be a function of $r=\s$, we conclude that $f$ is a constant. 

This means that the configuration (\ref{dependence}) is allowed just for particular metrics (\ref{Lif}) (as in \cite{Donos:2010tu}), namely 
those with $f$ constant, which occurs for instance when the internal manifold is $T^{1,1}$, whereas this configuration is forbidden for
the Sasaki-Einstein manifolds $Y^{p,q}$. It is rather curious that although we consider the string propagating just in the non-compact spacetime, 
the form of the internal manifold can determine physical aspects of the string propagation.

The equation of motion for $x=x(\s)$ is
\be 
\d_\s\left(\frac{g^2}{\sqrt{g^2\left((x')^2+(r')^2\right)}}\; \d_\s x\right)=0\;\Rightarrow 
\d_\s r=\pm V_{eff}\; \d_\s x\; ,
\ee
where
\be 
V_{eff}=\frac{1}{c_0}\sqrt{g^2(r)-c_0^2},
\ee
and $c_0$ is just an integration constant. We consider a  $\cap$--shaped string similar to the solution considered in the last section, 
namely a string which extends from $x=-\ell/2$ to $x=\ell/2$ and it reaches a maximum point $r_{max}$ in the bulk space. 

The boundary conditions for this configuration \cite{Nunez:2009da} imply that $\left.\frac{d r}{d x}\right|_{r\to 0}\to \infty$. 
In our case, we can easily see that this condition is satisfied since $\lim_{r\to 0}V_{eff}\to \infty$.

The turning point, i.e. the maximum point in the $r$ direction, is determined by the condition $\frac{d r}{d x}(r_{max})=0$, 
which gives
\be 
g^2(r_{max})-c_0^2=0\; \Rightarrow\; c_0^2=\frac{2L^4}{r_{max}^4}v_\xi - \frac{L^4 f}{r_{max}^2}v_\xi^2\; .
\ee
In order for $c_0$ to be real, we see that we need $v_\xi< 2/(f r^2_{max})$.

Finally, the distance between the string endpoints is
\be 
\ell_{q\bar{q}}(r_{max})=2 g(r_{max}) \int_0^{r_{max}} dr \frac{1}{\sqrt{g^2(r)-g^2(r_{max})}}\; ,
\ee
and if we define $w=r/r_{max}$ we find
\be 
\begin{split}
\ell_{q\bar{q}}(r_{max})&=\frac{2 r_{max}^3}{L^2\sqrt{v_\xi}} g(r_{max}) \int_0^1 dw \frac{w^2}{\sqrt{(fv_\xi r_{max}^2-2)w^4-fv_\xi r^2_{max}w^2+2}}\\
&\equiv \frac{2 r_{max}^3}{L^2\sqrt{v_\xi}} g(r_{max})\ {\cal I}(r_{max})\; .
\end{split}
\ee
In order to solve the integral, we write it as
\bse
\be 
{\cal I}(r_{max})=\int_0^1 dw \frac{w^2}{\sqrt{[(fv_\xi r_{max}^2-2)w^2-2](w^2-1)}}\; ,
\ee
and performing the substitution $w=\sin u $, we find the elliptic integral
\be 
{\cal I}(r_{max})=\frac{1}{\sqrt{2}}\int_0^{\pi/2} du \frac{ \sin^2 u }{\sqrt{1+\frac{(2-fv_\xi r^2_{max})} {2}\sin^2 u }}\; ,\label{integral}
\ee
\ese
with $(2-fv_\xi r^2_{max})>0$.

In terms of the complete elliptic integrals of first and second kind \cite{jeffrey2000table},
\bse
\begin{align}
\mathbf{K}(k)&=\int_0^{\pi/2}  \frac{ du }{\sqrt{1-k^2 \sin^2 u }}\\
\mathbf{E}(k)&=\int_0^{\pi/2} du \sqrt{1-k^2 \sin^2 u }
\end{align}
\ese
(The constant $k$ is called elliptic modulus and it can take any complex or real value \footnote{Generally in physics and engineering problems,
the modulus $k^2$ is parametrized in such that $k^2\in (0,1)$, but it is not our case. See \cite{byrd2013handbook} for details.}.) we can write (\ref{integral}) as
\be 
{\cal I}(r_{max})= \frac{\mathbf{K}(k)-\mathbf{E}(k)}{\sqrt{2} \ k^2}\; ,
\ee
where $k^2=(fv_\xi r^2_{max}-2)/2$. Then the distance between the string endpoints is given by
\be 
\begin{split}
\ell_{q\bar{q}}(r_{max})&=\frac{\sqrt{2}r_{max}^3}{L^2\sqrt{v_\xi}} g(r_{max}) \frac{\mathbf{K}(k)-\mathbf{E}(k)}{ \ k^2}\\
&=\frac{2\sqrt{2}}{\sqrt{f v_\xi}}\, \sqrt{-\frac{k^2+1}{k^2}\; } \left(\mathbf{K}(k)-\mathbf{E}(k)\right)\equiv 
\frac{2\sqrt{2}}{\sqrt{f v_\xi}}\, \L(-k^2)\; .
\end{split}
\ee

\begin{figure}[h]
 \centering
 \includegraphics[scale=0.5]{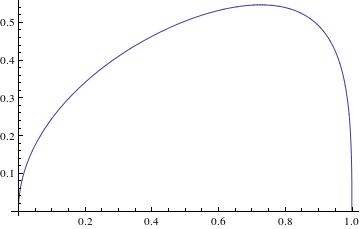}
  \caption{Graph of the function $\L(-k^2)$.}\label{figure01}
\end{figure}

Now observe that $-k^2=(2-fv_\xi r^2_{max})/2>0$, since $fv_\xi r^2_{max}<2$, therefore $k^2+1>0$, which implies that $v_{\xi}>0$. Therefore 
$v_\xi \in \left(0,\frac{2}{f r^2_{max}} \right)$, and $-k^2\in(0,1)$, see figure \ref{figure01}.

Finally, following the standard calculation \cite{Sonnenschein:1999if, Nunez:2009da}, we compute the observable that would correspond to the
energy of a $q\bar{q}$-pair (defined in relativistic gauge theories by introducing external quarks into the theory, and measuring their potential),
by subtracting from the string action the action of two `rods' that would fall from the end of the space to the boundary. 
The renormalized energy is obtained to be
\begin{align}
V_{q\bar{q}}(r_{max})&=2\int_0^{r_{max}} dr \frac{g^2(r)}{\sqrt{g^2(r)-g^2(r_{max})}} -2\int_0^{r_{max}} dr\; g(r)  \nonumber\\
&=\frac{\sqrt{2 v_\xi}L^2}{r_{max}}\int_0^1 dw \left[ \frac{(2w^{-2}-fv_\xi r^2_{max})}{\sqrt{\left( 1-\frac{(fv_\xi r^2_{max}-2)}{2}w^2 \right)\left(1-w^2 \right)}}
\right.\cr
&\left.-2 \frac{1}{w^2}\sqrt{\left( 1-\frac{fv_\xi r^2_{max}}{2}w^2 \right)}\right]\; \nonumber\\
& = \frac{\sqrt{2 v_\xi}L^2}{r_{max}} \left({\cal I}_{-2}(k, w^{-2}) + {\cal I}_0(k, w^0)-{\cal I}_{g}(k', w^{-2})\right)\; ,\label{potential}
\end{align}
where $k'=f v_\xi r^2_{max}/2$.

\begin{figure}[h]
        \centering
        \begin{subfigure}[b]{0.3\textwidth}
                \includegraphics[scale=0.3]{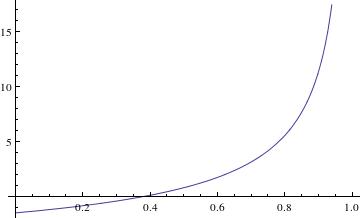}
                \caption{$a=0.1$}
        \end{subfigure}%
        ~ 
        \begin{subfigure}[b]{0.3\textwidth}
                \includegraphics[scale=0.3]{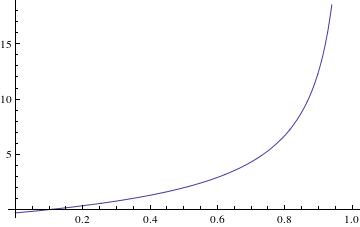}
                \caption{$a=1.0$}
        \end{subfigure}
        ~ 
        \begin{subfigure}[b]{0.3\textwidth}
                \includegraphics[scale=0.3]{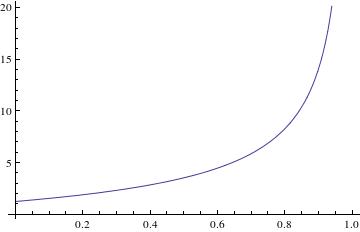}
                \caption{$a=1.9$}
        \end{subfigure}
        \caption{Graph of $\widehat{V}_{q\bar{q}}$ against $-k^2$, for three different values of $a$ and $n=0$.}\label{figure02}
\end{figure}

We can easily see that
\bse
\be 
{\cal I}_0(k, w^0)=\int_0^1 dw \frac{-fv_\xi r^2_{max}}{\sqrt{\left( 1-\frac{(fv_\xi r^2_{max}-2)}{2}w^2 \right)\left(1-w^2 \right)}}=-fv_\xi r^2_{max} \mathbf{K}(k)\; .
\ee
Consider the substitution $w=\sin u$, so that the second integral is
\be 
\begin{split}
{\cal I}_{-2}(k, w^{-2})&=2 \int_0^{\pi/2} du \frac{1}{\sin^2 u \sqrt{\left( 1-\frac{(fv_\xi r^2_{max}-2)}{2}\sin^2u \right)}}\\
&=2\left[ \mathbf{K}(k)-\mathbf{E}(k) \right]-2\left. \sqrt{\left( 1-\frac{(fv_\xi r^2_{max}-2)}{2}\sin^2u \right)}\cot u\right|_0^{\pi/2}\;,
\end{split}
\ee
and the third integral reads (considering the $arcsin$ modulo $2\pi n$)
\be 
\begin{split}
{\cal I}_{g}(k, w^{-2})&=2 \int_0^{\pi/2} du \frac{\cos u}{\sin^2 u}\sqrt{\left( 1-\frac{fv_\xi r^2_{max}}{2}\sin^2u \right)}\\
&=-\left[ 2\sqrt{\left( 1-\frac{fv_\xi r^2_{max}}{2}\sin^2u \right)}\csc u\right.\cr
&\left.\left.-\sqrt{2 fv_\xi r^2_{max}}\arcsin \left( \sqrt{\frac{fv_\xi r^2_{max}}{2}}
\sin u\right)\right]\right|_0^{\pi/2}\cr
&+2\pi n\sqrt{2f v_\xi r^2_{max}}\; .
\end{split}
\ee
\ese
The terms with $\arcsin(\, \cdots)$ and the terms in the upper limit $u\to \frac{\pi}{2}$ are constants, 
but we observe that we have two divergent terms for $u\to 0$, namely
\be 
\begin{split}
I=&-2\sqrt{\left( 1-\frac{fv_\xi r^2_{max}}{2}\sin^2u \right)}\csc u \\
II=&-2 \sqrt{\left( 1-\frac{(fv_\xi r^2_{max}-2)}{2}\sin^2u \right)}\cot u\; ,
\end{split}
\ee
and the difference in the equation (\ref{potential}) gives
\be 
\lim_{u\to 0}\left(-\sqrt{\left( 1-\frac{fv_\xi r^2_{max}}{2}\sin^2u \right)}\csc u+
 \sqrt{\left( 1-\frac{(fv_\xi r^2_{max}-2)}{2}\sin^2u \right)}\cot u \right)= 0\; .
\ee
All in all, if $n\in \mathbb{Z}$, the potential energy  $V_{q\bar{q}}$ is
\bse
\be 
\begin{split}
V_{q\bar{q}}(r_{max})&=\frac{\sqrt{2v_\xi}L^2}{r_{max}}\left[ (2-fv_\xi r^2_{max})\mathbf{K}(k)-2\mathbf{E}(k)-2\pi n \sqrt{2f v_\xi r^2_{max}}
  \phantom{\frac{1}{2}} \right. \\
& \hspace{2.0cm}\left.+ \sqrt{2 f v_\xi r_{max}^2 } \arcsin \left(\sqrt{\frac{f v_\xi r_{max}^2}{2}}\right) + 2\sqrt{1-\frac{f v_\xi r_{max}^2}{2}} \right]\; .
\end{split}
\ee
Since $v_\xi\in \left(0,\frac{2}{f r_{max}^2} \right)$, we write $v_\xi=\frac{a}{f r_{max}^2}$ with $a\in (0,2)$, such that
\be 
\begin{split}
\widehat{V}_{q\bar{q}}(r_{max})=\frac{1}{L^2 v_\xi \sqrt{f}}V_{q\bar{q}}(r_{max})&=\frac{1}{\sqrt{k^2+1}}\left[ -2\left(k^2\mathbf{K}(k)
+\mathbf{E}(k)\right)-2\pi n \sqrt{2 a}  \phantom{\frac{1}{2}}
 \right. \\
& \hspace{2.0cm}\left.+ \sqrt{2 a} \arcsin \left(\sqrt{\frac{a}{2}}\right) + 2\sqrt{1-\frac{a}{2}}\; \right]\; .
\end{split}
\ee
In the figure \ref{figure02}, we plot the graph for three different values of $a$.

Alternatively, we can write the energy as a function of the distance, $\ell_{q\bar{q}}$, as
\be
\begin{split} 
V_{q\bar{q}}(r_{max})=\frac{L^2 \sqrt{2 v_\xi}}{r_{max}} &\left[ -2\left(\frac{k^2 \sqrt{f v_\xi}}{2\sqrt{2}}\sqrt{\frac{-k^2}{1+k^2}} \ell_{q\bar{q}} +\frac{a}{2}\mathbf{E}(k)\right)\right. -2\pi n \sqrt{2a}\\ 
&\left.+ \sqrt{2a} \arcsin \left(\sqrt{\frac{a}{2}}\right) + 2\sqrt{1-\frac{a}{2}}\; \right] .
\end{split}
\ee
\ese
and from this last result we see that
\be 
\frac{d V_{q\bar{q}}}{d \ell_{q\bar{q}}}= L^2 \frac{(\sqrt{f} v_\xi)k^4/\sqrt{-k^2}}
{r_{max}(1+k^2)^{1/2}} >0\; , \quad \frac{d^2 V_{q\bar{q}}}{d \ell_{q\bar{q}}^2}=0\; .
\ee

It is important to notice that, although the potential energy $V_{q\bar{q}}$ exhibits a linear behaviour in relation to the distance $\ell_{q\bar{q}}$, 
similar to confining theories, we can not say that this theory is confining, since we have a maximum value for the distance $\ell_{q\bar{q}}$ in 
relation to the maximum distance $r_{max}$. Therefore, if we suppose that $\ell_{q\bar{q}}<\ell_{max}$, the potential $V_{q\bar{q}}$ is a 
bounded function of $\ell_{q\bar{q}}$. A similar phenomenon also happens in the calculations of Wilson 
loops at finite temperature \cite{Sonnenschein:1999if}. Moreover, as we said, it is not clear if the interpretation imported from the relativistic gauge 
theories still holds in this case.

\section{Discussion}

In this article we have studied nonabelian T-duality for non-relativistic holographic duals. In particular, using a NATD transformation we constructed 
novel examples of non-relativistic spaces with the interpretation of holographic duals, one for a conformal Galilean theory in massless type IIA, one
for a conformal Galilean theory in massive type IIA, and two for Lifshitz theories in type IIB, coming from NATD of spaces with $T^{1,1}$ and $Y^{p,q}$
internal spaces. 

In order to describe the field theories dual to the non-relativistic gravitational backgrounds, we have calculated the conserved charges of 
these backgrounds and we compared our results with those obtained in \cite{Lozano:2014ata}. 

We have also calculated the Wilson loop observables for the holographic dual spaces, though their true interpretation in the field theory remains 
to be seen, and it would be very interesting to understand. 
For the Wilson loops in gravity duals of conformal Galilean theories, we considered that the compact coordinate is constant and 
we found that the energy potential between quarks is always attractive. For the case of case of gravity dual of spaces with Lifshitz symmetry, 
we could not consider a constant compact coordinate, and we do not know the field theoretical interpretation for the string moving in this direction.
The Wilson loop that we found for this second class of spaces is proportional to the quark-antiquark distance, 
but the interpretation of this result is not clear.

It would be useful to characterize further the field theories dual to the non-relativistic backgrounds considered in this paper, by 
studying also other properties, like conductivity or shear viscosity.


{\bf Acknowledgements}. The research of HN is supported in part by CNPQ grant 301709/2013-0 and FAPESP grants 2013/14152-7
and 2014/18634-9. 
The work of TA is supported by CNPq grant 140588/2012-4. TA also would like to thank Eoin \'O Colg\'ain, \"Ozg\"ur Kelekci and Niall Macpherson for useful comments, and Guilherme Martins for help on condensed matter issues.

\appendix

\section{Non-Abelian T-Duality} \label{review}

In this Appendix we review non-Abelian T-duality, which is a transformation on a background that supports an $SU(2)$-structure. In principle we 
could follow the usual method employed in \cite{Sfetsos:2010uq,  Itsios:2013wd, Araujo:2015npa, Gevorgyan:2013xka, Macpherson:2014eza, Sfetsos:2014tza}, but here we review the strategy used by \cite{Kelekci:2014ima}, which has the advantage of giving a closed form for the fields 
in the RR sector \footnote{We thank to the authors of \cite{Kelekci:2014ima} for clarifications.}. Considering a spacetime metric and 
a Kalb-Ramond two-form given by
\begin{equation}
ds^2=G_{\mu\nu}(x) dx^\mu dx^\nu +2 G_{\mu i}(x) dx^\mu \tau_i+G_{ij}(x) \tau_i \tau_j \label{metric}
\end{equation}
\begin{equation}
B=\frac{1}{2}B_{\mu\nu} dx^\mu\wedge dx^\nu +B_{\mu i} dx^\mu \wedge \tau_i+ \frac{1}{2}B_{ij} \tau_i\wedge \tau_j, \label{kalbramond}
\end{equation}
in such a way that $\mu,\nu=1,\dots,7$ and all dependence on the $SU(2)$ angles $\theta, \psi, \phi$ is contained in the Maurer-Cartan 
forms $\tau_i$ for $SU(2)$, which satisfy $d\tau_i=\frac{1}{2}\epsilon_{ijk}\tau_j\wedge \tau_k$. Furthermore, in general this background 
has a non-trivial dilaton $\Phi=\Phi(x)$. 

If we define the field $Q$ by its components
\begin{equation}
\begin{split}
Q_{\mu\nu} &=G_{\mu\nu}+B_{\mu\nu}, \quad Q_{\mu i}=G_{\mu i}+B_{\mu i}\\
Q_{i \mu}&=G_{i \mu}+B_{i \mu},\quad \; E_{ij}=G_{ij}+B_{ij}\; ,
\end{split}
\end{equation}
one can show that the nonabelian T-dual background is given by
\begin{equation}
\begin{split}
\widehat{Q}_{\mu\nu} &=Q_{\mu\nu}-Q_{\mu i}M_{ij}^{-1}Q_{j\nu}, \quad \widehat{E}_{ij}=M_{ij}^{-1}\\
\widehat{Q}_{\mu i}&=Q_{ \mu j}M_{ji}^{-1}, \qquad \widehat{Q}_{i\mu}=-M_{ij}^{-1}Q_{ j\mu },
\end{split}
\end{equation}
where the matrix $M$ is defined by 
\begin{equation}
M_{ij}=E_{ij}+\a'\epsilon_{ijk} v_k\; ,
\end{equation}
and $\epsilon_{ijk}$ are the structure constants of the group $SU(2)$ and $v_i$ are Lagrange multipliers. Hereafter we absorb 
the factor of $\a'$ into $v_i$ and we present all the correct factors in the final answers. All in all, the dual fields are written as
\begin{subequations}
\begin{align}
d\hat{s}^2&=\widehat{G}_{\mu\nu}(x) dx^\mu dx^\nu +2 \widehat{G}_{\mu i}(x) dx^\mu dv^i+\widehat{G}_{ij}(x) dv^i dv^j \\
\widehat{B}&=\frac{1}{2}\widehat{B}_{\mu\nu} dx^\mu\wedge dx^\nu +\widehat{B}_{\mu i} dx^\mu \wedge dv^i+ \frac{1}{2}\widehat{B}_{ij} dv^i\wedge dv^j\; ,
\end{align}
and the one-loop contribution to the dilaton is given by
\begin{equation}
\hat{\phi}=\phi-\frac{1}{2}\ln \left(\frac{\Delta}{\a'^3}\right)\; ,
\end{equation}
\end{subequations}
where $\Delta=\det M$. Besides the spectator fields, the dual theory depends on $\theta, \psi, \phi, v^i$, meaning that we have too many 
degrees of freedom and we need to impose a gauge fixing in order to remove three of these variables. 

It is convenient to write the metric as
\begin{equation}
ds_{10}^2=ds_7^2+\sum_{i=1}^3e^{2C_i}(\tau_i+A^i)^2\; , \label{spmetric}
\end{equation}
where $A^i$ are $SU(2)$-valued gauge fields and $C_i$ are scalars. Moreover, we define the vielbeins $\{\mathfrak{e}^\mu, \mathfrak{e}^i\}$, such that
\begin{equation}
\begin{split}
\mathfrak{e}^\mu &\; \Rightarrow\; ds_7^{2}=g_{\mu\nu}dx^\mu dx^\nu=\sum_{\mu=0}^6(\mathfrak{e}^\mu)^2\\
\mathfrak{e}^i=e^{C_i}(\tau_i+A^i)&\; \Rightarrow\; \sum_{i=1}^3e^{2C_i}(\tau_i+A^i)^2=\sum_{i=1}^3(\mathfrak{e}^i)^2,
\end{split}
\end{equation}
implying that the components of the metric (\ref{metric}) are
\begin{equation}
G_{\mu\nu}=g_{\mu\nu}+\sum_{i=1}^3 e^{2C_i}A^i_\mu A^i_\nu\; ,\quad G_{\mu i}=e^{2C_i} A_\mu^i\; ,\quad G_{ij}=e^{2C_i}\delta_{ij}\; .
\end{equation}
In the same way, it is useful to write the Kalb-Ramond as
\begin{equation}
B=\frac{1}{2}b_{\mu\nu} dx^\mu\wedge dx^\nu+(\b_{i}+db_i) \wedge \tau_i+ \frac{1}{2}\epsilon_{ijk}b_{k} \tau_i\wedge \tau_j\; ,
\end{equation}
and the components of (\ref{kalbramond}) are
\begin{equation}
B_{\mu\nu}=b_{\mu\nu}\; ,\quad B_{\mu i}=\beta_{\mu i}+\partial_\mu b_i\; ,\quad B_{ij}=\epsilon_{ijk}b_k\; .
\end{equation}
We next write the inverse of the matrix $M_{ij}$,
\begin{equation}
M_{ij}^{-1}=\frac{1}{\Delta}
\begin{pmatrix}
e^{2(C_2+C_3)}+z_1^2\; & z_1 z_2- e^{2C_3} z_3\; & z_1 z_3+e^{2C_2}z_2\\
z_1 z_2+e^{2C_3} z_3\; & e^{2(C_1+C_3)}+z_2^2\; & z_2z_3-e^{2C_1} z_1\\
z_1 z_3-e^{2C_2}z_2\; & z_2z_3+e^{2C_1} z_1 \; & e^{2(C_1+C_2)}+z_3^2
\end{pmatrix}\; ,
\end{equation}
where $\Delta= e^{2(C_1+C_2+C_3)}+e^{2 C_1}z_1^2+e^{2 C_2}z_2^2+e^{2 C_3}z_3^2$ and $z_i=\a' v_i+b_i$. In fact, it is easy to see 
the general formula for the components of $M^{-1}$ is
\begin{equation}
M_{ij}^{-1}=\frac{1}{\Delta}(z_i z_j+\delta_{ij}e^{2(C_1+C_2+C_3)}e^{-2C_i}-\epsilon_{ijk} e^{2 C_k}z_k)\; .
\end{equation}
Using all these equations, the authors of \cite{Kelekci:2014ima} were able to find a closed form for the dual metric and Kalb-Ramond field,
\bse
\begin{equation}
\begin{split}
d\hat{s}^2&=ds^2_7+\frac{1}{\Delta}\left[(z_1Dz_1+z_2Dz_2+z_3Dz_3)^2+e^{2(C_2+C_3)}Dz_1^2\right.\\
&\hspace{2.2cm}\left.+e^{2(C_1+C_3)}Dz_2^2+e^{2(C_1+C_2)}Dz_3^2\right]
\end{split}
\end{equation}
\begin{equation}
\begin{split}
&\widehat{B}=\frac{1}{2}B_{\mu\nu}dx^\mu \wedge dx^\nu-\frac{1}{\Delta}(e^{2C_1}z_1 Dz_2 \wedge Dz_3+e^{2C_2}z_2 Dz_3
 \wedge Dz_1+e^{2C_3}z_3 Dz_1 \wedge Dz_2)\\
&-Dz_1\wedge A_1-Dz_2\wedge A_2-Dz_3\wedge A_3-z_1A_2\wedge A_3-z_2A_3\wedge A_1-z_3A_1\wedge A_2\; 
\end{split}
\end{equation}
where 
\begin{equation}
Dz_i=dz_i+\b_i-\epsilon_{ijk} z_j A^k \; .
\end{equation}
\ese
For the RR sector the authors of \cite{Kelekci:2014ima} have shown explicit closed forms for the dual backgrounds. 
Considering first a (massive) type IIA sector with fields given by
\bse
\begin{align}
F_0&= m\label{rra}\\
F_2&= G_2+J_1^i\wedge (\tau_i+A^i)+\frac{1}{2}\epsilon_{ijk}K_0^i(\tau_j+A^j)\wedge (\tau_k+A^k)\\
F_4&=G_4+L_3^i \wedge (\tau_i+A^i)+\frac{1}{2}\epsilon_{ijk}M_2^i\wedge(\tau_j+A^j)\wedge (\tau_k+A^k)\nonumber\\
&+N_1\wedge (\tau_1+A^1)\wedge (\tau_2+A^2)\wedge (\tau_3+A^3)\; ,
\end{align}
\ese
one can find the dual type IIB RR fields as
\bse
\begin{align}
\a'^{3/2}\widehat{F}_1&= m z_i e^{C_i} \hat{\mathfrak{e}}^i- z_iJ^i_1-K_0^i e^{C_i} \hat{\mathfrak{e}}^i
+ \epsilon_{ijk}K_0^iz_j e^{-C_k} \hat{\mathfrak{e}}^k+ N_1\\
\a'^{3/2}\widehat{F}_3&= m e^{C_1+C_2+C_3} \hat{\mathfrak{e}}^1\wedge \hat{\mathfrak{e}}^2 \wedge \hat{\mathfrak{e}}^3
+e^{C_1+C_2+C_3}  \ast_7 G_4+G_2\wedge z_i e^{C_i}\hat{\mathfrak{e}}^i\nonumber\\
&-\frac{1}{2}\epsilon_{ijk}J_1^i\wedge e^{C_j+C_k}\hat{\mathfrak{e}}^j\wedge \hat{\mathfrak{e}}^k
+J_1^i\wedge e^{-C_i} \hat{\mathfrak{e}}^i \wedge z_j e^{C_j} \hat{\mathfrak{e}}^j\nonumber\\
&+z_i K_0^i e^{2C_i}e^{-C_1-C_2-C_3} \hat{\mathfrak{e}}^1\wedge\hat{\mathfrak{e}}^2\wedge \hat{\mathfrak{e}}^3\\
&- N_1\wedge \frac{1}{2}\epsilon_{ijk}z_ie^{-C_b-C_c} \hat{\mathfrak{e}}^j\wedge \hat{\mathfrak{e}}^k
- z_i L_3^i-M_2^i\wedge e^{C_i} \hat{\mathfrak{e}}^i+ \epsilon_{ijk} M_2^i z_j\wedge e^{-C_k} \hat{\mathfrak{e}}^k\nonumber\\
\a'^{3/2} \widehat{F}_5&=(1+\ast)\left[G_4\wedge z_i e^{C_i} \hat{\mathfrak{e}}^ai+ e^{C_1+C_2+C_3}G_2\wedge \hat{\mathfrak{e}}^1
\wedge \hat{\mathfrak{e}}^2 \wedge \hat{\mathfrak{e}}^3\right.\cr
&-\frac{1}{2}\epsilon_{ijk}L_3^i \wedge e^{C_j+C_k} \hat{\mathfrak{e}}^j\wedge 
\hat{\mathfrak{e}}^k\nonumber\\
&+\left. L_3^i\wedge e^{-C_i} \hat{\mathfrak{e}}^i\wedge z_j e^{C_j} \hat{\mathfrak{e}}^j+z_i M_2^{i} e^{2C_i} \wedge 
e^{-C_1-C_2-C_3} \hat{\mathfrak{e}}^1\wedge \hat{\mathfrak{e}}^2\wedge \hat{\mathfrak{e}}^3\right]\label{rre}\; .
\end{align}
\ese
Reversely, starting from a type IIB solution, with RR-fields given by
\bse
\begin{align}
F_1&=G_1\\
F_3&=G_3+X_2^i\wedge (\tau_i+A^i)+\frac{1}{2}\epsilon_{ijk}Y_1^i\wedge  (\tau_j+A^j)\wedge  (\tau_k+A^k)\nonumber\\
&+m  (\tau_1+A^1)\wedge  (\tau_2+A^2)\wedge  (\tau_3+A^3)\\
F_5&=(1+\ast)\left[Z_4^i\wedge  (\tau_i+A^i)+G_2\wedge  (\tau_1+A^1)\wedge  (\tau_2+A^2) \wedge  (\tau_3+A^3)   \right]\; ,
\end{align}
we have the dual fields in IIA supergravity \footnote{Observe that, compared with \cite{Kelekci:2014ima}, we have some 
different signs in the dual RR-fields. We thank Eoin \'O Colg\'ain for letting us know about it.}
\begin{align}
\widehat{F}_0&=-m\\
\widehat{F}_2&=-e^{C_a}z_a G_1\wedge \hat{\mathfrak{e}}^a+z_a X_2^a+Y_1^a\wedge(e^{C_a}\hat{\mathfrak{e}}^a)
-\epsilon^{abc}Y_1^a\wedge (z_b e^{-C_c}\hat{\mathfrak{e}}^c)\nonumber\\
&+\frac{m}{2}e^{-C_a-C_b}\epsilon^{abc}z_c \hat{\mathfrak{e}}^a\wedge \hat{\mathfrak{e}}^b-G_2\\
\widehat{F}_4&=e^{C_1+C_2+C_3}\ast_7 G_3-z_aZ^a_4 - e^{C_1+C_2+C_3} \ast_7 Z_4^a\wedge (e^{-C_a}\hat{\mathfrak{e}}^a)
-\epsilon^{abc}e^{-C_a}\ast_7 Z_4^a\wedge z_be^{C_b} \hat{\mathfrak{e}}^c\nonumber\\
&-e^{C_a}z_a G_3\wedge \hat{\mathfrak{e}}^a+\frac{1}{2}G_2\wedge(\epsilon^{abc}z_a e^{-C_b-C_c}\hat{\mathfrak{e}}^b\wedge 
\hat{\mathfrak{e}}^c)+\frac{1}{2}\epsilon^{abc}X_2^a\wedge (e^{C_b+C_c}\hat{\mathfrak{e}}^b\wedge \hat{\mathfrak{e}}^c)\nonumber\\
&-X_2^a\wedge (e^{-C_a}\hat{\mathfrak{e}}^a)\wedge (z_be^{C_b} \hat{\mathfrak{e}}^b)-e^{C_1+C_2+C_3}G_1
\wedge \hat{\mathfrak{e}}^1\wedge \hat{\mathfrak{e}}^2 \wedge \hat{\mathfrak{e}}^3\nonumber\\
&-z_a e^{2C_a}Y_1^a\wedge (e^{-C_1-C_2-C_3} \hat{\mathfrak{e}}^1\wedge \hat{\mathfrak{e}}^2 \wedge \hat{\mathfrak{e}}^3) 
\end{align}
\ese
where the dual vielbeins are defined to be 
\begin{equation}
\hat{\mathfrak{e}}^{i}= e^{C_i}\Delta^{-1}
\left\{
z_i z_j Dz_j+e^{2\sum_{i\neq j}C_j}Dz_i+\epsilon_{ijk}z_j e^{2C_j}Dz_k
\right\}
 \; .
\end{equation}

\newpage

\bibliographystyle{utphys}
\bibliography{NATDNonRel}{}

\providecommand{\href}[2]{#2}\begingroup\raggedright\begin{thebibliography}{10}

\bibitem{Maldacena:1997re}
J.~M. Maldacena, ``{The Large N limit of superconformal field theories and
  supergravity},'' \href{http://dx.doi.org/10.1023/A:1026654312961}{{\em
  Int.J.Theor.Phys.} {\bf 38} (1999)  1113--1133},
\href{http://arxiv.org/abs/hep-th/9711200}{{\tt arXiv:hep-th/9711200
  [hep-th]}}.

\bibitem{Gubser:1998bc}
S.~Gubser, I.~R. Klebanov, and A.~M. Polyakov, ``{Gauge theory correlators from
  noncritical string theory},''
  \href{http://dx.doi.org/10.1016/S0370-2693(98)00377-3}{{\em Phys.Lett.} {\bf
  B428} (1998)  105--114},
\href{http://arxiv.org/abs/hep-th/9802109}{{\tt arXiv:hep-th/9802109
  [hep-th]}}.

\bibitem{Witten:1998qj}
E.~Witten, ``{Anti-de Sitter space and holography},'' {\em
  Adv.Theor.Math.Phys.} {\bf 2} (1998)  253--291,
\href{http://arxiv.org/abs/hep-th/9802150}{{\tt arXiv:hep-th/9802150
  [hep-th]}}.

\bibitem{Aharony:1999ti}
O.~Aharony, S.~S. Gubser, J.~M. Maldacena, H.~Ooguri, and Y.~Oz, ``{Large N
  field theories, string theory and gravity},''
  \href{http://dx.doi.org/10.1016/S0370-1573(99)00083-6}{{\em Phys.Rept.} {\bf
  323} (2000)  183--386},
\href{http://arxiv.org/abs/hep-th/9905111}{{\tt arXiv:hep-th/9905111
  [hep-th]}}.

\bibitem{anderson2008basic}
P.~Anderson, {\em Basic Notions Of Condensed Matter Physics}.
\newblock Advanced Books Classics Series. Westview Press, 2008.

\bibitem{sachdev2011quantum}
S.~Sachdev, {\em Quantum Phase Transitions}.
\newblock Troisi{\`e}me Cycle de la Physique. Cambridge University Press, 2011.

\bibitem{Nishida:2007pj}
Y.~Nishida and D.~T. Son, ``{Nonrelativistic conformal field theories},''
  \href{http://dx.doi.org/10.1103/PhysRevD.76.086004}{{\em Phys.Rev.} {\bf D76}
  (2007)  086004},
\href{http://arxiv.org/abs/0706.3746}{{\tt arXiv:0706.3746 [hep-th]}}.

\bibitem{Son:2008ye}
D.~Son, ``{Toward an AdS/cold atoms correspondence: A Geometric realization of
  the Schrodinger symmetry},''
  \href{http://dx.doi.org/10.1103/PhysRevD.78.046003}{{\em Phys.Rev.} {\bf D78}
  (2008)  046003},
\href{http://arxiv.org/abs/0804.3972}{{\tt arXiv:0804.3972 [hep-th]}}.

\bibitem{Balasubramanian:2008dm}
K.~Balasubramanian and J.~McGreevy, ``{Gravity duals for non-relativistic
  CFTs},'' \href{http://dx.doi.org/10.1103/PhysRevLett.101.061601}{{\em
  Phys.Rev.Lett.} {\bf 101} (2008)  061601},
\href{http://arxiv.org/abs/0804.4053}{{\tt arXiv:0804.4053 [hep-th]}}.

\bibitem{Herzog:2008wg}
C.~P. Herzog, M.~Rangamani, and S.~F. Ross, ``{Heating up Galilean
  holography},'' \href{http://dx.doi.org/10.1088/1126-6708/2008/11/080}{{\em
  JHEP} {\bf 0811} (2008)  080},
\href{http://arxiv.org/abs/0807.1099}{{\tt arXiv:0807.1099 [hep-th]}}.

\bibitem{Maldacena:2008wh}
J.~Maldacena, D.~Martelli, and Y.~Tachikawa, ``{Comments on string theory
  backgrounds with non-relativistic conformal symmetry},''
  \href{http://dx.doi.org/10.1088/1126-6708/2008/10/072}{{\em JHEP} {\bf 0810}
  (2008)  072},
\href{http://arxiv.org/abs/0807.1100}{{\tt arXiv:0807.1100 [hep-th]}}.

\bibitem{Adams:2008wt}
A.~Adams, K.~Balasubramanian, and J.~McGreevy, ``{Hot Spacetimes for Cold
  Atoms},'' \href{http://dx.doi.org/10.1088/1126-6708/2008/11/059}{{\em JHEP}
  {\bf 0811} (2008)  059},
\href{http://arxiv.org/abs/0807.1111}{{\tt arXiv:0807.1111 [hep-th]}}.

\bibitem{Hartnoll:2009sz}
S.~A. Hartnoll, ``{Lectures on holographic methods for condensed matter
  physics},'' \href{http://dx.doi.org/10.1088/0264-9381/26/22/224002}{{\em
  Class.Quant.Grav.} {\bf 26} (2009)  224002},
\href{http://arxiv.org/abs/0903.3246}{{\tt arXiv:0903.3246 [hep-th]}}.

\bibitem{Kachru:2008yh}
S.~Kachru, X.~Liu, and M.~Mulligan, ``{Gravity duals of Lifshitz-like fixed
  points},'' \href{http://dx.doi.org/10.1103/PhysRevD.78.106005}{{\em
  Phys.Rev.} {\bf D78} (2008)  106005},
\href{http://arxiv.org/abs/0808.1725}{{\tt arXiv:0808.1725 [hep-th]}}.

\bibitem{Balasubramanian:2010uk}
K.~Balasubramanian and K.~Narayan, ``{Lifshitz spacetimes from AdS null and
  cosmological solutions},''
  \href{http://dx.doi.org/10.1007/JHEP08(2010)014}{{\em JHEP} {\bf 1008} (2010)
   014},
\href{http://arxiv.org/abs/1005.3291}{{\tt arXiv:1005.3291 [hep-th]}}.

\bibitem{Donos:2010tu}
A.~Donos and J.~P. Gauntlett, ``{Lifshitz Solutions of D=10 and D=11
  supergravity},'' \href{http://dx.doi.org/10.1007/JHEP12(2010)002}{{\em JHEP}
  {\bf 1012} (2010)  002},
\href{http://arxiv.org/abs/1008.2062}{{\tt arXiv:1008.2062 [hep-th]}}.

\bibitem{Bobev:2009mw}
N.~Bobev, A.~Kundu, and K.~Pilch, ``{Supersymmetric IIB Solutions with
  Schrodinger Symmetry},''
  \href{http://dx.doi.org/10.1088/1126-6708/2009/07/107}{{\em JHEP} {\bf 0907}
  (2009)  107},
\href{http://arxiv.org/abs/0905.0673}{{\tt arXiv:0905.0673 [hep-th]}}.

\bibitem{Donos:2009en}
A.~Donos and J.~P. Gauntlett, ``{Supersymmetric solutions for non-relativistic
  holography},'' \href{http://dx.doi.org/10.1088/1126-6708/2009/03/138}{{\em
  JHEP} {\bf 0903} (2009)  138},
\href{http://arxiv.org/abs/0901.0818}{{\tt arXiv:0901.0818 [hep-th]}}.

\bibitem{Donos:2009xc}
A.~Donos and J.~P. Gauntlett, ``{Solutions of type IIB and D=11 supergravity
  with Schrodinger(z) symmetry},''
  \href{http://dx.doi.org/10.1088/1126-6708/2009/07/042}{{\em JHEP} {\bf 0907}
  (2009)  042},
\href{http://arxiv.org/abs/0905.1098}{{\tt arXiv:0905.1098 [hep-th]}}.

\bibitem{Ooguri:2009cv}
H.~Ooguri and C.-S. Park, ``{Supersymmetric non-relativistic geometries in
  M-theory},'' \href{http://dx.doi.org/10.1016/j.nuclphysb.2009.08.021}{{\em
  Nucl.Phys.} {\bf B824} (2010)  136--153},
\href{http://arxiv.org/abs/0905.1954}{{\tt arXiv:0905.1954 [hep-th]}}.

\bibitem{Donos:2009zf}
A.~Donos and J.~P. Gauntlett, ``{Schrodinger invariant solutions of type IIB
  with enhanced supersymmetry},''
  \href{http://dx.doi.org/10.1088/1126-6708/2009/10/073}{{\em JHEP} {\bf 0910}
  (2009)  073},
\href{http://arxiv.org/abs/0907.1761}{{\tt arXiv:0907.1761 [hep-th]}}.

\bibitem{Singh:2010rt}
H.~Singh, ``{Galilean type IIA backgrounds and a map},''
  \href{http://dx.doi.org/10.1142/S0217732311035791}{{\em Mod.Phys.Lett.} {\bf
  A26} (2011)  1443--1451},
\href{http://arxiv.org/abs/1007.0866}{{\tt arXiv:1007.0866 [hep-th]}}.

\bibitem{Jeong:2009aa}
J.~Jeong, H.-C. Kim, S.~Lee, E.~\'O~Colg\'ain, and H.~Yavartanoo,
  ``{Schr\"odinger invariant solutions of M-theory with Enhanced
  Supersymmetry},'' \href{http://dx.doi.org/10.1007/JHEP03(2010)034}{{\em JHEP}
  {\bf 1003} (2010)  034},
\href{http://arxiv.org/abs/0911.5281}{{\tt arXiv:0911.5281 [hep-th]}}.

\bibitem{Gregory:2010gx}
R.~Gregory, S.~L. Parameswaran, G.~Tasinato, and I.~Zavala, ``{Lifshitz
  solutions in supergravity and string theory},''
  \href{http://dx.doi.org/10.1007/JHEP12(2010)047}{{\em JHEP} {\bf 12} (2010)
  047},
\href{http://arxiv.org/abs/1009.3445}{{\tt arXiv:1009.3445 [hep-th]}}.

\bibitem{Ko:2015rha}
S.~M. Ko, C.~Melby-Thompson, R.~Meyer, and J.-H. Park, ``{Dynamics of
  Perturbations in Double Field Theory and Non-Relativistic String Theory},''
\href{http://arxiv.org/abs/1508.01121}{{\tt arXiv:1508.01121 [hep-th]}}.

\bibitem{Gauntlett:2004hs}
J.~P. Gauntlett, D.~Martelli, J.~Sparks, and D.~Waldram, ``{Supersymmetric AdS
  backgrounds in string and M-theory},''
\href{http://arxiv.org/abs/hep-th/0411194}{{\tt arXiv:hep-th/0411194
  [hep-th]}}.

\bibitem{Gran:2005wu}
U.~Gran, G.~Papadopoulos, and D.~Roest, ``{Systematics of M-theory spinorial
  geometry},'' \href{http://dx.doi.org/10.1088/0264-9381/22/13/013}{{\em
  Class.Quant.Grav.} {\bf 22} (2005)  2701--2744},
\href{http://arxiv.org/abs/hep-th/0503046}{{\tt arXiv:hep-th/0503046
  [hep-th]}}.

\bibitem{Lazaroiu:2013fg}
C.~I. Lazaroiu and E.-M. Babalic, ``{Geometric algebra and M-theory
  compactifications},'' {\em Rom.J.Phys.} {\bf 58} (2013)  5--6,
\href{http://arxiv.org/abs/1301.5094}{{\tt arXiv:1301.5094 [hep-th]}}.

\bibitem{Buscher:1987qj}
T.~Buscher, ``{Path Integral Derivation of Quantum Duality in Nonlinear Sigma
  Models},''
\href{http://dx.doi.org/10.1016/0370-2693(88)90602-8}{{\em Phys.Lett.} {\bf
  B201} (1988)  466}.

\bibitem{Rocek:1991ps}
M.~Rocek and E.~P. Verlinde, ``{Duality, quotients, and currents},''
  \href{http://dx.doi.org/10.1016/0550-3213(92)90269-H}{{\em Nucl.Phys.} {\bf
  B373} (1992)  630--646},
\href{http://arxiv.org/abs/hep-th/9110053}{{\tt arXiv:hep-th/9110053
  [hep-th]}}.

\bibitem{Giveon:1991jj}
A.~Giveon and M.~Rocek, ``{Generalized duality in curved string backgrounds},''
  \href{http://dx.doi.org/10.1016/0550-3213(92)90518-G}{{\em Nucl.Phys.} {\bf
  B380} (1992)  128--146},
\href{http://arxiv.org/abs/hep-th/9112070}{{\tt arXiv:hep-th/9112070
  [hep-th]}}.

\bibitem{Alvarez:1993qi}
E.~\'Alvarez, L.~\'Alvarez-Gaum\'e, J.~Barbon, and Y.~Lozano, ``{Some global
  aspects of duality in string theory},''
  \href{http://dx.doi.org/10.1016/0550-3213(94)90067-1}{{\em Nucl.Phys.} {\bf
  B415} (1994)  71--100},
\href{http://arxiv.org/abs/hep-th/9309039}{{\tt arXiv:hep-th/9309039
  [hep-th]}}.

\bibitem{Alvarez:1994dn}
E.~\'Alvarez, L.~\'Alvarez-Gaum\'e, and Y.~Lozano, ``{An Introduction to T
  duality in string theory},''
  \href{http://dx.doi.org/10.1016/0920-5632(95)00429-D}{{\em
  Nucl.Phys.Proc.Suppl.} {\bf 41} (1995)  1--20},
\href{http://arxiv.org/abs/hep-th/9410237}{{\tt arXiv:hep-th/9410237
  [hep-th]}}.

\bibitem{delaOssa:1992vc}
X.~C. de~la Ossa and F.~Quevedo, ``{Duality symmetries from nonAbelian
  isometries in string theory},''
  \href{http://dx.doi.org/10.1016/0550-3213(93)90041-M}{{\em Nucl.Phys.} {\bf
  B403} (1993)  377--394},
\href{http://arxiv.org/abs/hep-th/9210021}{{\tt arXiv:hep-th/9210021
  [hep-th]}}.

\bibitem{Giveon:1993ai}
A.~Giveon and M.~Rocek, ``{On nonAbelian duality},''
  \href{http://dx.doi.org/10.1016/0550-3213(94)90230-5}{{\em Nucl.Phys.} {\bf
  B421} (1994)  173--190},
\href{http://arxiv.org/abs/hep-th/9308154}{{\tt arXiv:hep-th/9308154
  [hep-th]}}.

\bibitem{Gasperini:1993nz}
M.~Gasperini, R.~Ricci, and G.~Veneziano, ``{A Problem with nonAbelian
  duality?},'' \href{http://dx.doi.org/10.1016/0370-2693(93)91748-C}{{\em
  Phys.Lett.} {\bf B319} (1993)  438--444},
\href{http://arxiv.org/abs/hep-th/9308112}{{\tt arXiv:hep-th/9308112
  [hep-th]}}.

\bibitem{Alvarez:1994np}
E.~\'Alvarez, L.~\'Alvarez-Gaum\'e, and Y.~Lozano, ``{On nonAbelian duality},''
  \href{http://dx.doi.org/10.1016/0550-3213(94)90093-0}{{\em Nucl.Phys.} {\bf
  B424} (1994)  155--183},
\href{http://arxiv.org/abs/hep-th/9403155}{{\tt arXiv:hep-th/9403155
  [hep-th]}}.

\bibitem{Sfetsos:2010uq}
K.~Sfetsos and D.~C. Thompson, ``{On non-abelian T-dual geometries with Ramond
  fluxes},'' \href{http://dx.doi.org/10.1016/j.nuclphysb.2010.12.013}{{\em
  Nucl.Phys.} {\bf B846} (2011)  21--42},
\href{http://arxiv.org/abs/1012.1320}{{\tt arXiv:1012.1320 [hep-th]}}.

\bibitem{Lozano:2011kb}
Y.~Lozano, E.~\'O~Colg\'ain, K.~Sfetsos, and D.~C. Thompson, ``{Non-abelian
  T-duality, Ramond Fields and Coset Geometries},''
  \href{http://dx.doi.org/10.1007/JHEP06(2011)106}{{\em JHEP} {\bf 1106} (2011)
   106},
\href{http://arxiv.org/abs/1104.5196}{{\tt arXiv:1104.5196 [hep-th]}}.

\bibitem{Itsios:2013wd}
G.~Itsios, C.~N\'u\~nez, K.~Sfetsos, and D.~C. Thompson, ``{Non-Abelian
  T-duality and the AdS/CFT correspondence: new N=1 backgrounds},''
  \href{http://dx.doi.org/10.1016/j.nuclphysb.2013.04.004}{{\em Nucl.Phys.}
  {\bf B873} (2013)  1--64},
\href{http://arxiv.org/abs/1301.6755}{{\tt arXiv:1301.6755 [hep-th]}}.

\bibitem{Kelekci:2014ima}
{\" O}.~Kelekci, Y.~Lozano, N.~T. Macpherson, and E.~\'O~Colg\'ain,
  ``{Supersymmetry and non-Abelian T-duality in type II supergravity},''
  \href{http://dx.doi.org/10.1088/0264-9381/32/3/035014}{{\em
  Class.Quant.Grav.} {\bf 32} (2015) no.~3, 035014},
\href{http://arxiv.org/abs/1409.7406}{{\tt arXiv:1409.7406 [hep-th]}}.

\bibitem{Lozano:2012au}
Y.~Lozano, E.~\'O~Colg\'ain, D.~Rodr\'{i}guez-G\'{o}mez, and K.~Sfetsos,
  ``{Supersymmetric $AdS_6$ via T Duality},''
  \href{http://dx.doi.org/10.1103/PhysRevLett.110.231601}{{\em Phys.Rev.Lett.}
  {\bf 110} (2013) no.~23, 231601},
\href{http://arxiv.org/abs/1212.1043}{{\tt arXiv:1212.1043 [hep-th]}}.

\bibitem{Itsios:2012zv}
G.~Itsios, C.~N\'u\~nez, K.~Sfetsos, and D.~C. Thompson, ``{On Non-Abelian
  T-Duality and new N=1 backgrounds},''
  \href{http://dx.doi.org/10.1016/j.physletb.2013.03.033}{{\em Phys.Lett.} {\bf
  B721} (2013)  342--346},
\href{http://arxiv.org/abs/1212.4840}{{\tt arXiv:1212.4840}}.

\bibitem{Lozano:2013oma}
Y.~Lozano, E.~\'O~Colg\'ain, and D.~Rodr\'{i}guez-G\'{o}mez, ``{Hints of 5d
  Fixed Point Theories from Non-Abelian T-duality},''
  \href{http://dx.doi.org/10.1007/JHEP05(2014)009}{{\em JHEP} {\bf 1405} (2014)
   009},
\href{http://arxiv.org/abs/1311.4842}{{\tt arXiv:1311.4842 [hep-th]}}.

\bibitem{Caceres:2014uoa}
E.~Caceres, N.~T. Macpherson, and C.~N\'u\~nez, ``{New Type IIB Backgrounds and
  Aspects of Their Field Theory Duals},''
  \href{http://dx.doi.org/10.1007/JHEP08(2014)107}{{\em JHEP} {\bf 1408} (2014)
   107},
\href{http://arxiv.org/abs/1402.3294}{{\tt arXiv:1402.3294 [hep-th]}}.

\bibitem{Lozano:2014ata}
Y.~Lozano and N.~T. Macpherson, ``{A new AdS$_{4}$/CFT$_{3}$ dual with extended
  SUSY and a spectral flow},''
  \href{http://dx.doi.org/10.1007/JHEP11(2014)115}{{\em JHEP} {\bf 1411} (2014)
   115},
\href{http://arxiv.org/abs/1408.0912}{{\tt arXiv:1408.0912 [hep-th]}}.

\bibitem{Araujo:2015npa}
T.~R. Araujo and H.~Nastase, ``{$\mathcal{N}=1$ SUSY backgrounds with an AdS
  factor from non-Abelian T duality},''
  \href{http://dx.doi.org/10.1103/PhysRevD.91.126015}{{\em Phys. Rev.} {\bf
  D91} (2015) no.~12, 126015},
\href{http://arxiv.org/abs/1503.00553}{{\tt arXiv:1503.00553 [hep-th]}}.

\bibitem{Bea:2015fja}
Y.~Bea, J.~D. Edelstein, G.~Itsios, K.~S. Kooner, C.~N\'u\~nez, {\em et al.},
  ``{Compactifications of the Klebanov-Witten CFT and new AdS$_{3}$
  backgrounds},'' \href{http://dx.doi.org/10.1007/JHEP05(2015)062}{{\em JHEP}
  {\bf 1505} (2015)  062},
\href{http://arxiv.org/abs/1503.07527}{{\tt arXiv:1503.07527 [hep-th]}}.

\bibitem{Macpherson:2013zba}
N.~T. Macpherson, ``{Non-Abelian T-duality, $G_2$-structure rotation and
  holographic duals of $N=1$ Chern-Simons theories},''
  \href{http://dx.doi.org/10.1007/JHEP11(2013)137}{{\em JHEP} {\bf 11} (2013)
  137},
\href{http://arxiv.org/abs/1310.1609}{{\tt arXiv:1310.1609 [hep-th]}}.

\bibitem{Barranco:2013fza}
A.~Barranco, J.~Gaillard, N.~T. Macpherson, C.~N\'u\~nez, and D.~C. Thompson,
  ``{G-structures and Flavouring non-Abelian T-duality},''
  \href{http://dx.doi.org/10.1007/JHEP08(2013)018}{{\em JHEP} {\bf 08} (2013)
  018},
\href{http://arxiv.org/abs/1305.7229}{{\tt arXiv:1305.7229 [hep-th]}}.

\bibitem{Gaillard:2013vsa}
J.~Gaillard, N.~T. Macpherson, C.~N\'u\~nez, and D.~C. Thompson, ``{Dualising
  the Baryonic Branch: Dynamic SU(2) and confining backgrounds in IIA},''
  \href{http://dx.doi.org/10.1016/j.nuclphysb.2014.05.004}{{\em Nucl. Phys.}
  {\bf B884} (2014)  696--740},
\href{http://arxiv.org/abs/1312.4945}{{\tt arXiv:1312.4945 [hep-th]}}.

\bibitem{Kooner:2014cqa}
K.~S. Kooner and S.~Zacar\'ias, ``{Non-Abelian T-Dualizing the Resolved
  Conifold with Regular and Fractional D3-Branes},''
  \href{http://dx.doi.org/10.1007/JHEP08(2015)143}{{\em JHEP} {\bf 08} (2015)
  143},
\href{http://arxiv.org/abs/1411.7433}{{\tt arXiv:1411.7433 [hep-th]}}.

\bibitem{Lozano:2015bra}
Y.~Lozano, N.~T. Macpherson, J.~Montero, and E.~\'O~Colg\'ain, ``{New $AdS_3
  \times S^2$ T-duals with $ \mathcal{N}=\left(0,4\right) $ supersymmetry},''
  \href{http://dx.doi.org/10.1007/JHEP08(2015)121}{{\em JHEP} {\bf 08} (2015)
  121},
\href{http://arxiv.org/abs/1507.02659}{{\tt arXiv:1507.02659 [hep-th]}}.

\bibitem{Lozano:2015cra}
Y.~Lozano, N.~T. Macpherson, and J.~Montero, ``{A $\mathcal{N}=2$
  Supersymmetric $AdS_4$ Solution in M-theory with Purely Magnetic Flux},''
\href{http://arxiv.org/abs/1507.02660}{{\tt arXiv:1507.02660 [hep-th]}}.

\bibitem{Aharony:2008ug}
O.~Aharony, O.~Bergman, D.~L. Jafferis, and J.~Maldacena, ``{N=6 superconformal
  Chern-Simons-matter theories, M2-branes and their gravity duals},''
  \href{http://dx.doi.org/10.1088/1126-6708/2008/10/091}{{\em JHEP} {\bf 0810}
  (2008)  091},
\href{http://arxiv.org/abs/0806.1218}{{\tt arXiv:0806.1218 [hep-th]}}.

\bibitem{Nishioka:2008gz}
T.~Nishioka and T.~Takayanagi, ``{On Type IIA Penrose Limit and N=6
  Chern-Simons Theories},''
  \href{http://dx.doi.org/10.1088/1126-6708/2008/08/001}{{\em JHEP} {\bf 0808}
  (2008)  001},
\href{http://arxiv.org/abs/0806.3391}{{\tt arXiv:0806.3391 [hep-th]}}.

\bibitem{Cvetic:2000yp}
M.~Cvetic, H.~Lu, and C.~Pope, ``{Consistent warped space Kaluza-Klein
  reductions, half maximal gauged supergravities and CP**n constructions},''
  \href{http://dx.doi.org/10.1016/S0550-3213(00)00708-2}{{\em Nucl.Phys.} {\bf
  B597} (2001)  172--196},
\href{http://arxiv.org/abs/hep-th/0007109}{{\tt arXiv:hep-th/0007109
  [hep-th]}}.

\bibitem{Gevorgyan:2013xka}
E.~Gevorgyan and G.~Sarkissian, ``{Defects, Non-abelian T-duality, and the
  Fourier-Mukai transform of the Ramond-Ramond fields},''
  \href{http://dx.doi.org/10.1007/JHEP03(2014)035}{{\em JHEP} {\bf 1403} (2014)
   035},
\href{http://arxiv.org/abs/1310.1264}{{\tt arXiv:1310.1264 [hep-th]}}.

\bibitem{Macpherson:2014eza}
N.~T. Macpherson, C.~N\'{u}\~nez, L.~A. Pando~Zayas, V.~G.~J. Rodgers, and
  C.~A. Whiting, ``{Type IIB supergravity solutions with AdS$_{5}$ from Abelian
  and non-Abelian T dualities},''
  \href{http://dx.doi.org/10.1007/JHEP02(2015)040}{{\em JHEP} {\bf 1502} (2015)
   040},
\href{http://arxiv.org/abs/1410.2650}{{\tt arXiv:1410.2650 [hep-th]}}.

\bibitem{Gauntlett:2004yd}
J.~P. Gauntlett, D.~Martelli, J.~Sparks, and D.~Waldram, ``{Sasaki-Einstein
  metrics on S**2 x S**3},''
  \href{http://dx.doi.org/10.4310/ATMP.2004.v8.n4.a3}{{\em
  Adv.Theor.Math.Phys.} {\bf 8} (2004)  711--734},
\href{http://arxiv.org/abs/hep-th/0403002}{{\tt arXiv:hep-th/0403002
  [hep-th]}}.

\bibitem{Hull:1998vy}
C.~M. Hull, ``{Massive string theories from M theory and F theory},''
  \href{http://dx.doi.org/10.1088/1126-6708/1998/11/027}{{\em JHEP} {\bf 11}
  (1998)  027},
\href{http://arxiv.org/abs/hep-th/9811021}{{\tt arXiv:hep-th/9811021
  [hep-th]}}.

\bibitem{smilga2001lectures}
A.~Smilga, {\em Lectures on Quantum Chromodynamics}.
\newblock World Scientific, 2001.

\bibitem{nastase2015introduction}
H.~Nastase, {\em Introduction to the AdS/CFT Correspondence}.
\newblock Cambridge University Press, 2015.

\bibitem{erdmenger2015}
J.~Erdmenger and M.~Ammon, {\em Gauge/gravity duality. Foundations and
  applications}.
\newblock Cambridge University Press, 2015.

\bibitem{Nastase:2007kj}
H.~Nastase, ``{Introduction to AdS-CFT},''
\href{http://arxiv.org/abs/0712.0689}{{\tt arXiv:0712.0689 [hep-th]}}.

\bibitem{Sonnenschein:1999if}
J.~Sonnenschein, ``{What does the string / gauge correspondence teach us about
  Wilson loops?},''
\href{http://arxiv.org/abs/hep-th/0003032}{{\tt arXiv:hep-th/0003032
  [hep-th]}}.

\bibitem{Nunez:2009da}
C.~N\'u\~nez, M.~Piai, and A.~Rago, ``{Wilson Loops in string duals of Walking
  and Flavored Systems},''
  \href{http://dx.doi.org/10.1103/PhysRevD.81.086001}{{\em Phys.Rev.} {\bf D81}
  (2010)  086001},
\href{http://arxiv.org/abs/0909.0748}{{\tt arXiv:0909.0748 [hep-th]}}.

\bibitem{Araujo:2014kda}
T.~R. Araujo and H.~Nastase, ``{Comments on the T-dual of the gravity dual of
  D5-branes on S$^{3}$},''
  \href{http://dx.doi.org/10.1007/JHEP04(2015)081}{{\em JHEP} {\bf 1504} (2015)
   081},
\href{http://arxiv.org/abs/1409.7333}{{\tt arXiv:1409.7333 [hep-th]}}.

\bibitem{Danielsson:2009gi}
U.~H. Danielsson and L.~Thorlacius, ``{Black holes in asymptotically Lifshitz
  spacetime},'' \href{http://dx.doi.org/10.1088/1126-6708/2009/03/070}{{\em
  JHEP} {\bf 0903} (2009)  070},
\href{http://arxiv.org/abs/0812.5088}{{\tt arXiv:0812.5088 [hep-th]}}.

\bibitem{Kluson:2009vy}
J.~Kluson, ``{Open String in Non-Relativistic Background},''
  \href{http://dx.doi.org/10.1103/PhysRevD.81.106006}{{\em Phys.Rev.} {\bf D81}
  (2010)  106006},
\href{http://arxiv.org/abs/0912.4587}{{\tt arXiv:0912.4587 [hep-th]}}.

\bibitem{Fadafan:2009an}
K.~B. Fadafan, ``{Drag force in asymptotically Lifshitz spacetimes},''
\href{http://arxiv.org/abs/0912.4873}{{\tt arXiv:0912.4873 [hep-th]}}.

\bibitem{Siahaan:2011sw}
H.~M. Siahaan, ``{On non-relativistic $Q\bar Q$ potential via Wilson loop in
  Galilean space-time},''
  \href{http://dx.doi.org/10.1142/S0217732311036255}{{\em Mod.Phys.Lett.} {\bf
  A26} (2011)  1719--1724},
\href{http://arxiv.org/abs/1106.5008}{{\tt arXiv:1106.5008 [hep-th]}}.

\bibitem{Fadafan:2015iwa}
K.~B. Fadafan and F.~Saiedi, ``{On Holographic Non-relativistic Schwinger
  Effect},''
\href{http://arxiv.org/abs/1504.02432}{{\tt arXiv:1504.02432 [hep-th]}}.

\bibitem{Bachas:1985xs}
C.~Bachas, ``{Convexity of the Quarkonium Potential},''
\href{http://dx.doi.org/10.1103/PhysRevD.33.2723}{{\em Phys.Rev.} {\bf D33}
  (1986)  2723}.

\bibitem{Arias:2009me}
R.~E. Arias and G.~A. Silva, ``{Wilson loops stability in the gauge/string
  correspondence},'' \href{http://dx.doi.org/10.1007/JHEP01(2010)023}{{\em
  JHEP} {\bf 1001} (2010)  023},
\href{http://arxiv.org/abs/0911.0662}{{\tt arXiv:0911.0662 [hep-th]}}.

\bibitem{jeffrey2000table}
A.~Jeffrey and D.~Zwillinger, {\em Table of Integrals, Series, and Products}.
\newblock Elsevier Science, 2000.

\bibitem{byrd2013handbook}
P.~Byrd and M.~Friedman, {\em Handbook of Elliptic Integrals for Engineers and
  Physicists}.
\newblock Grundlehren der mathematischen Wissenschaften. Springer Berlin
  Heidelberg, 2013.

\bibitem{Sfetsos:2014tza}
K.~Sfetsos and D.~C. Thompson, ``{New ${\cal N} = 1$ supersymmetric $AdS_5$
  backgrounds in Type IIA supergravity},''
  \href{http://dx.doi.org/10.1007/JHEP11(2014)006}{{\em JHEP} {\bf 1411} (2014)
   006},
\href{http://arxiv.org/abs/1408.6545}{{\tt arXiv:1408.6545 [hep-th]}}.

\end{thebibliography}\endgroup

 
\end{document}